\documentclass[prb,twocolumn,floatfix,a4paper]{revtex4}

\usepackage{dcolumn}
\usepackage{graphicx,subfigure}
\usepackage{amsmath,amssymb}

\DeclareMathOperator{\trace}{Tr}
\newcommand{\SIGMA}{\boldsymbol{\sigma}}

\newcommand{\deriv}{\mathrm{d}}

\newcommand{\cf}{\textit{cf.}{ }}
\newcommand{\etal}{\textit{et al.}{ }}
\newcommand{\ie}{\textit{i.e.}{ }}

\newcommand{\abinitio}{\textit{ab-initio}{ }}

\newcolumntype{.}[1]{D{.}{.}{#1}}
\begin{document}

\title{First principles study of the solubility of Zr in Al}

\author{Emmanuel \surname{Clouet}}
\email{emmanuel.clouet@cea.fr}
\altaffiliation{Current address: Service de Recherches de M\'etallurgie Physique, CEA/Saclay,
91191 Gif-sur-Yvette, France}
\author{J.~M. \surname{Sanchez}}
\affiliation{Texas Materials Institute, The University of Texas at Austin, Austin, Texas 78712}
\author{C. \surname{Sigli}}
\affiliation{Pechiney Centre de Recherches de Voreppe, B.P.~27,
38341  Voreppe~cedex, France}

\date{\today}

%%%%%%%%%%%%%%%%%%%%%%%%%%%%%%%%%%%%%%%%%
\begin{abstract}
The experimental solubility limit of Zr in Al is well-known. Al$_3$Zr has a
stable structure DO$_{23}$ and a metastable one L1$_2$. Consequently there is
a metastable solubility limit for which only few experimental data are
available. The purpose of this study is to obtain by \abinitio calculations
the solubility limit of Zr in Al for the stable as well as the metastable
phase diagrams.
The formation energies of several ordered compounds Al$_x$Zr$_{(1-x)}$,
all based on an fcc underlying lattice, were calculated using the FP-LMTO
(Full Potential Linear Muffin Tin Orbital) method.
Taking into account all the relaxations allowed by the symmetry, we found
the DO$_{23}$ structure to be the stable one for Al$_3$Zr.
This set of results was then used with the cluster expansion in order to fit
a generalized Ising model through the inverse method of Connolly-Williams.
Different ways to consider volume relaxations were examined.
This allowed us to calculate in the Bragg-Williams approximation the
configurational free energy at finite temperature.
According to the previous FP-LMTO calculations the free energy due to
electronic excitations can be neglected.
For the vibrational free energy of ordered structures we compared results
obtained from a calculation of the elastic constants used with the Debye model
and results obtained from a calculation of the phonon spectrum.
All these different steps lead to a calculation of the solubility limit of Zr
in Al which is found to be lower than the experimental one.
The solubility limit in the metastable phase diagram is calculated in the same
way and can thus be compared to the stable one.
\end{abstract}

\maketitle

%%%%%%%%%%%%%%%%%%%%%%%%%%%%%%%%%%%%%

\section{Introduction}

The development of methods based on the density functional theory\cite{HOH64,KOH65}
and of the computer power has allowed to conceive calculations of phase diagram
from first principles\cite{DUCASTELLE,DEF94} as an alternative to laboratory experimentations.
Traditionally, only substitutional effects were considered, which was good enough to reproduce
the topology of most phase diagrams.
So as to obtain a more quantitative agreement with experimental data, it has been shown more recently that
electronic excitations\cite{WOL95} as well as lattice vibrations\cite{GAR94,OZO01} could play an important
part in the relative stability of different phases.
We chose to test the ability of these first principles methods to predict the solubility limit of Zr in an
aluminium solid solution this part of the Al-Zr phase diagram being interesting because of the presence
of a metastable phase.

\begin{figure}[!hbtp]
\includegraphics[height=0.45\textwidth]{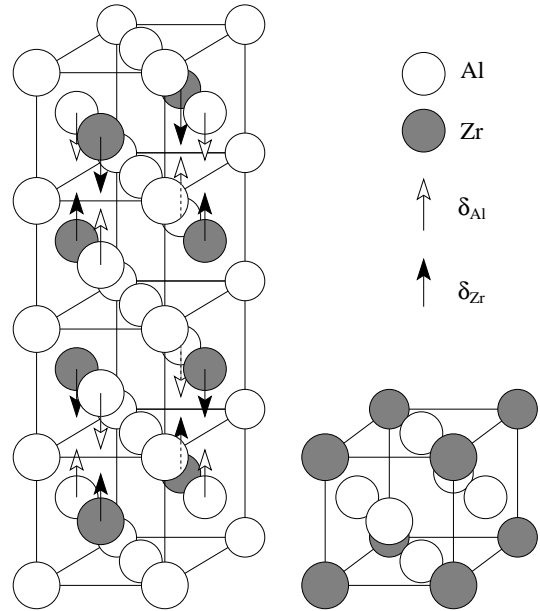}
\caption{Definition of the structures DO$_{23}$ (left) and L1$_2$ (right).}
\label{al3zr_fig}
\end{figure}

The Al-richest intermediate phase of the Al-Zr phase diagram\footnote{The whole phase diagram
has been extensively reviewed by Murray \etal\cite{MUR92}}
is Al$_3$Zr. This compound has the DO$_{23}$
structure which is body-centered tetragonal with eight atoms
per unit cell, some of these atoms being allowed by symmetry to move along the main
axis of the unit cell (Fig. \ref{al3zr_fig}).
It is stable up to 1580$\pm$10$^{\circ}$C.

The solubility limit of Zr in Al~(fcc) is really low, the maximum solubility being 0.083~at.\%
at the peritectic reaction, Liquid+ZrAl$_3$ $\longleftrightarrow$ (Al).
The solubilities of Zr in both liquid and solid were definitively
determined by Fink and Willey \cite{FIN39} and the assessed
phase diagram is based on their data.
The solid solubility was determined from resistivity data and checked by metallography.
Solid solubilities were also reported by Glazov \etal \cite{GLA59},
Drits \etal \cite{DRI68} (solubilities determined by means of
microstructural analysis and electric resistivity measurements),
and Kuznetsov \etal \cite{KUZ83} (determined from resistivity,
microhardness and lattice constant measurements as well as metallography).
The solubilities reported in these last studies are higher than the ones
of the assessed phase diagram.

Supersaturated solid solution of Zr in (Al) containing as much as 3~at.\% Zr
can be prepared by rapid solidification.
A coherent metastable phase Al$_3$Zr precipitates from the supersaturated
solution\cite{RYU69}. This phase has the structure L1$_2$ which is simple cubic with 4
atoms per unit cell (Fig. \ref{al3zr_fig}).
This metastable phase can also form from the melt as a primary phase during
rapid solidification\cite{NES72,NES77}: Al$_3$Zr acts as  nuclei for solidification of (Al)
and Zr can thus work as a grain refiner of Al.
This metastable phase is also responsible for the effectiveness of Zr to control
recrystallization in Al-based alloys: it leads to a more uniform
distribution of fine precipitates that pins grains and subgrains boundaries.
Moreover, this phase is really stable against coarsening and redissolution,
all this making it highly desirable.
%Moreover, because of its cubic symmetry, it is expected to be more ductile
%than the DO$_{23}$ structure\cite{VAR91}. This added to its high melting temperature,
%its low density, and its good oxidation resistance makes it highly desirable.
As few experimental data are available for this phase, it is hard to
fit a thermodynamic model for it.
In such a case, a first principles calculation of the phase diagram should
allow us to obtain properties that are not available experimentally.

In order to assess the metastable phase diagram, we studied with the same
tools and approximations the L1$_2$ and DO$_{23}$ phases:
this allows us to check first the agreement between the stable phase diagram obtained
and experimental data, and then to compare it to the metastable phase diagram.

In a first part, we study the stability of ordered compounds based on
an fcc underlying lattice in the Al-Zr system.
The energies of different structures were calculated using an
ab-initio method, the full-potential-linear-muffin-tin-orbital (FP-LMTO).
The equilibrium parameters, like the volume, shape of the unit cell, or
positions of atoms, were obtained. For stable structures they
can be compared to experimental data.

Using this whole set of calculations we applied the cluster expansion to deduce
the energy of any structure based on the same underlying lattice in the
Al-Zr system, examining carefully the way to include volume relaxations.

At finite temperature, the electronic excitations,
the vibrational free energy, and the configurational entropy
have to be taken into account.
At the end of this part, we are able to obtain a thermodynamic model
written in the same way as in a Calphad approach and to calculate the
corresponding solubility limits.

%%%%%%%%%%%%%%%%%%%%%%%%%%%%%%%%%%%%%%%%%%%%%%

\section{Ground states of ordered compounds}

Formation energies were calculated at absolute zero temperature for 26
compounds in the Al-Zr binary system, all based on an fcc
lattice. Calculations were carried out using a full-potential
linear-muffin-tin-orbital (FP-LMTO) method \cite{AND75,MET88,MET89} in the
version developed by Methfessel and Van Schilfgaarde \cite{MET93}.
The basis used contained 22 energy independent muffin-tin-orbitals (MTO)
per Al and Zr site:
three $\kappa$ values for the orbitals s and p and two $\kappa$ values for
the orbitals d where the corresponding kinetic energies were
$\kappa^2=0.01$ Ry (spd), $1.0$ Ry (spd), and $2.3$ Ry (sp).
A second panel with a basis composed of 22 energy independent MTO with
the same kinetic energies was used to make a correct treatment of the
4p semicore states of Zr.
The same uniform mesh of points was used to make the integrations in the
Brillouin zone for valence and semicore states. The number of divisions
along reciprocal vectors was chosen big enough to ensure a convergence of
the total energy of the order of 0.1 mRy/atom for each structure.
The radii of the muffin-tin spheres were chosen to have a compactness of
47.6\% for Al sites and 58.4\% for Zr sites.
Inside the muffin-tin spheres, the potential is expanded in spherical
harmonics up to $l=6$ and in the interstitial region spherical Hankel
functions of kinetic energies $\kappa^2=1$ Ry and $3.0$ Ry were fitted up
to $l=6$.
The calculations were performed in the local density approximation (LDA)
\cite{HOH64,KOH65} and the parametrization used was the one of von
Barth-Hedin \cite{VON72}. Jomard \etal \cite{JOM98} showed that
generalized-gradient corrections have to be included in order to obtain a
correct description of the stability of the different phases of pure Zr,
but as we were interested only in the Al-rich part of the phase diagram
we did not include these gradient corrections.

Ground state energies at equilibrium $E_0$, equilibrium volumes per atom
$V_0$, and bulk moduli $B$ were obtained by fitting the Rose equation of state
\cite{ROS84} to the energies calculated for different volumes around the
minimum
\begin{equation}
E(r)=E_0\left(1+\frac{r-r_0}{\delta}\right)
\exp\left(-\frac{r-r_0}{\delta}\right),
\label{Rose}
\end{equation}
where $r$ is the Wigner-Seitz radius associated to the atomic volume $V$
and $\delta$ is related to the bulk modulus $B$ through the relation
\begin{equation}
B=\frac{-E_0}{12\pi r_0 \delta^2}.
\end{equation}

For the different compounds, the energies were optimized with respect to the
volume and all other degrees of freedom allowed by the symmetry, like the shape
of the unit cell or some atomic positions.
The equilibrium volumes $V_0$, bulk moduli $B$, and formation energies
$E^{form}$ relative to the fcc phases of both pure Al and Zr are presented in
table \ref{AlZr_relaxed_lda}. We note that all the formation energies are
negative, and they thus characterize Al-Zr as an ordering system.

\begin{table*}[!hbt]

\caption{Equilibrium volume $V_0$, bulk modulus $B$,
and formation energy $E^{form}$ for relaxed Al-Zr
compounds calculated with LDA.}
\label{AlZr_relaxed_lda}

\begin{ruledtabular}
\begin{tabular}{lll.{-1}.{-1}.{-1}}
& \multicolumn{1}{c}{Pearson}   & \multicolumn{1}{c}{Structure} &
\multicolumn{1}{c}{$V_0$}       & \multicolumn{1}{c}{$B$}       &
\multicolumn{1}{c}{$E^{form}$}  \\
& \multicolumn{1}{c}{symbol}    & \multicolumn{1}{c}{type}      &
\multicolumn{1}{c}{(\AA$^3$/atom)}& \multicolumn{1}{c}{(GPa)}   &
\multicolumn{1}{c}{(mRy/atom)\footnote{Reference phases are Al(fcc) and Zr(fcc).}} \\

\hline
Al (fcc)             & cF4   & Cu           & 15.82     & 80.78   & 0.    \\
Al$_{31}$Zr          & cP32  & ?           & 15.99     & 82.56   & -3.04 \\
Al$_{15}$Zr          & cI32  & ?            & 16.10     & 84.27   & -6.99 \\
Al$_8$Zr             & tI18  & V$_4$Zn$_5$  & 16.25     & 86.24   & -9.77 \\
Al$_7$Zr (D1)        & cF32  & Ca$_7$Ge     & 16.30     & 87.32   & -14.33\\
Al$_4$Zr (D1$_{\textrm{a}}$)& tI10 & MoNi$_4$ & 16.58   & 92.10   & -21.12\\
Al$_3$Zr (L1$_2$)    & cP4   & Cu$_3$Au     & 16.12     & 99.59   & -39.00\\
Al$_3$Zr (DO$_{22}$) & tI8   & Al$_3$Ti     & 16.60     & 99.65   & -39.04\\
Al$_3$Zr (DO$_{23}$) & tI16  & Al$_3$Zr     & 16.35     & 100.05  & -40.72\\
Al$_2$Zr ($\alpha$)  & hP3   & CdI$_2$      & 18.01     & 87.16   & -11.73\\
Al$_2$Zr ($\beta$)   & tI6   & MoSi$_2$     & 17.13     & 96.40   & -26.19\\
Al$_2$Zr ($\gamma$)  & oI6   & MoPt$_2$     & 17.15     & 96.51   & -26.08\\
AlZr (L1$_0$)        & tP4   & AuCu         & 18.15     & 103.33  & -37.07\\
AlZr (L1$_1$)        & hR32  & CuPt         & 19.04     & 93.29   & -16.50\\
AlZr (CH40)          & tI8   & NbP          & 18.52     & 100.48  & -33.56\\
AlZr (D4)            & cF32  & ?            & 18.49     & 92.58   & -14.78\\
AlZr (Z2)            & tP8   & ?            & 18.63     & 99.70   & -21.03\\
Zr$_2$Al ($\alpha$)  & hP3   & CdI$_2$      & 20.38     & 98.10   & -10.72\\
Zr$_2$Al ($\beta$)   & tI6   & MoSi$_2$     & 19.53     & 104.84  & -24.78\\
Zr$_2$Al ($\gamma$)  & oI6   & MoPt$_2$     & 19.44     & 104.05  & -26.36\\
Zr$_3$Al (L1$_2$)    & cP4   & Cu$_3$Au     & 19.71     & 107.67  & -27.11\\
Zr$_3$Al (DO$_{22}$) & tI8   & Al$_3$Ti     & 19.88     & 105.14  & -23.49\\
Zr$_3$Al (DO$_{23}$) & tI16  & Al$_3$Zr     & 19.80     & 102.77  & -25.18\\
Zr$_4$Al (D1$_{\textrm{a}}$)& tI10 & MoNi$_4$ & 20.31   & 99.85   & -16.30\\
Zr$_7$Al (D1)        & cF32  & Ca$_7$Ge     & 20.93     & 101.66  & -7.96 \\
Zr (fcc)             & cF4   & Cu           & 21.70     & 98.74   & 0.    \\

\end{tabular}
\end{ruledtabular}
\end{table*}

We examined more closely the stability of the phases L1$_2$, DO$_{22}$,
and DO$_{23}$ of Al$_3$Zr according to relaxations.
L1$_2$ being cubic, its energy was optimized with respect to atomic
volume only,
whereas in the tetragonal DO$_{22}$ phase optimization was performed
additionally with respect to the $c/a$ ratio
and in the tetragonal DO$_{23}$ phase to the $c/a$ ratio and to the
atomic displacements $\delta_{\textrm{Al}}$ and $\delta_{\textrm{Zr}}$
($\delta_{\textrm{Al}}$ and $\delta_{\textrm{Zr}}$ are defined on Fig. \ref{al3zr_fig}).
% To compare to experimental data, we calculated the energy of formation
% of these compounds relative to the fcc phase of pure Al and
% the hcp phase of pure Zr.
% The cohesive energy of the hcp phase of Zr at equilibrium was obtained after
% optimization with respect to the atomic volume and the $c/a$ ratio.
% We obtained $V_0=21.92$ \AA$^3$/atom and $c/a=1.616$,
% to be compared to the experimental values \cite{PEARSON}
% $V_0=23.28$ \AA$^3$/atom and $c/a=1.593$.

\begin{table*}[!hbt]

\caption{Calculated volumes $V_0$, $c'/a$ ratios
($c'=c/2$ for DO$_{22}$ phase and $c'=c/4$ for DO$_{23}$ phase),
atomic displacements (normalized by $c'$),
and ground state energies relative to the DO$_{23}$ phase for Al$_3$Zr
compared to previous calculations and experimental data.}
\label{Al3Zr_calcul}

\begin{ruledtabular}
\begin{tabular}{llccccc}

&Method& $V_0$         & $c'/a$     & Atomic       & $\Delta E$\\
&& (\AA$^3$/atom)&          &  displacements & (mRy/atom)\\
\hline
L1$_2$    & Present work & 16.12 &    &   & 1.71 \\
& FP-LMTO \cite{AMA95}   &       &    &   & 0.64 \\
& VASP \cite{COL01}      & 17.4  &    &   & 2.3  \\
& Experiments \cite{DES91}& 17.14 &   &   & 1.69 \\

DO$_{22}$ & Present work & 16.60 & 1.141 && 1.68 \\
& FP-LMTO \cite{AMA95}   &       &       && 2.63 \\
& VASP \cite{COL01}      & 17.7  & 1.141 && 1.9  \\

DO$_{23}$ & Present work & 16.35 & 1.087 & $\delta_{\textrm{Al}}=-0.0021$ &0.\\
&&&&$\delta_{\textrm{Zr}}=-0.0273$     \\
& FP-LMTO \cite{AMA95}   & 16.28 & 1.09  & $\delta_{\textrm{Al}}=+0.003$  &0.\\
&&&&$\delta_{\textrm{Zr}}=-0.026$     \\
& VASP \cite{COL01}      & 17.5  & 1.079 & $\delta_{\textrm{Al}}=+0.0003$ &0.\\
&&&&$\delta_{\textrm{Zr}}=-0.0101$     \\

&Experiments\cite{AMA95} & 17.25 & 1.0775 &$\delta_{\textrm{Al}}=+0.0004$  & \\
&&&&$\delta_{\textrm{Zr}}=-0.0272$\\

\end{tabular}
\end{ruledtabular}
\end{table*}

Our results for Al$_3$Zr agree really well with the experimental ones (Table \ref{Al3Zr_calcul}).
The equilibrium volumes obtained  for the L1$_2$ and DO$_{23}$ phases are
lower than the experimental ones, but this is a known feature of LDA.
This can be improved by adding gradient corrections: Colinet and
Pasturel\cite{COL01} using the generalized gradient approximation instead of LDA
found a better agreement with experimental data for these equilibrium volumes.
After relaxing all the degrees of freedom, we see that DO$_{23}$ is
the stable phase.
As been shown previously by Amador \etal \cite{AMA95} using the FP-LMTO
technique too, it is not enough to consider only the relaxation of the shape
of the unit cell ($c/a$ ratio) of the phase DO$_{23}$ to stabilize it,
the atomic displacements $\delta_{\textrm{Al}}$ and $\delta_{\textrm{Zr}}$
allowed by the symmetry have to be relaxed too,
otherwise L1$_2$ still has a lower energy.
This was confirmed by Colinet and Pasturel \cite{COL01} with
calculations in the pseudopotential method, and we observed such a
behaviour too.
The values obtained after relaxation for these displacements are close
to those measured by neutron diffraction by Amador \etal \cite{AMA95}:
the sign of $\delta_{\textrm{Al}}$ is wrong but this relative displacement
is too close from 0 to be really significant.
The enthalpy of transformation from the L1$_2$ to the DO$_{23}$ structure
was measured by Desch \etal \cite{DES91}. The experimental value
($\Delta H=-1.69$~mRy/atom) agrees really well with the value obtained from
ours calculations ($\Delta H=-1.72$~mRy/atom), which was not the case of
previous calculations.

For Zr$_3$Al, we found the phase L1$_2$ to have the lowest formation energy
compared to the two other structures we investigated.
This is in agreement with the experimental fact that L1$_2$ is the stable
phase of Zr$_3$Al.
For other compositions, the experimental stable structures are not based
on an fcc underlying lattice, and therefore no direct comparison can be made
with our calculations.

%%%%%%%%%%%%%%%%%%%%%%%%%%%%%%%%%%%%%%%%

\section{Cluster expansion of the formation energy}
%%%%%%%%%%

The FP-LMTO method only allows one to calculate the energy of perfectly ordered
systems which contain a few atoms per unit cell.
Disordered or partially ordered systems can be modeled by super-cells, but this
requires a too large computational time.
Moreover, to compute the free energy of these systems, one needs to calculate
the energy of every configuration. This cannot
be directly done with \abinitio calculations and a cluster expansion has to
be used.
That is why in the following we will directly use the FP-LMTO calculations only
for the perfectly ordered compounds Al$_3$Zr in the structures L1$_2$ and DO$_{23}$,
and for the solid solution Al-Zr we will make a cluster expansion.

\subsection{Formalism}

Considering a binary crystal of N sites on a rigid lattice,
its configuration can be described through an Ising model
by the vector $\SIGMA =\{\sigma_1, \sigma_2, \ldots, \sigma_N\}$
where the pseudo-spin configuration variable $\sigma_p$ equals
$\pm1$ if an A or B atom occupies the site $p$.
Any structure is then defined by its density matrix $\rho^s$,
$\rho^s(\boldsymbol{\sigma})$ being the probability of finding the
structure $s$ in the configuration $\SIGMA$.

To any cluster of $n$ lattice points $\alpha=\{i_1,i_2,\ldots,i_n\}$
is associated the following multisite correlation function
\begin{equation}
\zeta_{\alpha}^s
= \trace  \rho^s \prod_{i\in\alpha}\sigma_i
= \frac{1}{2^N} \sum_{\SIGMA} \rho^s(\SIGMA) \prod_{i\in\alpha}\sigma_i
\label{correlation},
\end{equation}
where the sum has to be performed over the $2^N$ possible configurations of
the lattice.

Clusters related by a translation or a symmetry operation of the point
group of the structure have the same correlation functions.
Denoting $D_{\alpha}$ the number of such equivalent clusters per
lattice site, or degeneracy, any configurational function $f^s$
can be expanded in the form \cite{SAN84}
\begin{equation}
f^s= \sum_{\alpha} D_{\alpha} f_{\alpha} \zeta_{\alpha}^s
\label{ce},
\end{equation}
where the sum has to be performed over all non equivalent clusters and
the coefficients $f_{\alpha}$ are independent of the structure.

The usefulness of the expansion (\ref{ce}) rests on the fast convergence of
these coefficients with the size of the clusters,
\ie with the number of points included in the cluster as well as
the maximal distance between points inside the cluster.
This allows one to truncate the sum using only a finite number of clusters.
Knowing the value of the function $f^s$ for a finite set of structures,
the coefficients $f_{\alpha}$ can then be obtained using the inverse method
proposed by Connolly and Williams \cite{CON83}, \ie by a matrix inversion
if the number of structures is the same as the number of clusters used in
the truncated expansion.
Here we used more structures than clusters and obtained the coefficients by
a least square fit. We can thus check the convergence of the
expansion by its ability to reproduce $f^s$.

Rather than doing the fit directly on the configurational function $f^s$, we
did it on the excess function associated which is defined as
\begin{equation}
\Delta f^s = f^s - \frac{1+\zeta_1^s}{2}f^A - \frac{1-\zeta_1^s}{2}f^B ,
\end{equation}
where $\zeta_1^s$ is the point correlation
and $f^A$ and $f^B$ are the values of the function $f^s$ for a lattice
occupied by respectively only atoms A ($\zeta_1=1$) and atoms B
($\zeta_1=-1$).
In the case of the energy, this excess function is nothing else but the
formation energy.

Using the expansion (\ref{ce}) we obtained
\begin{equation}
\Delta f^s = \sum_{\alpha, |\alpha|\ge2} D_{\alpha} f_{\alpha}
\left[ \zeta_{\alpha}^s - \frac{1+(-1)^{|\alpha|}}{2}
- \zeta_1^s  \frac{1-(-1)^{|\alpha|}}{2} \right]
\label{ce_excess} ,
\end{equation}
where $|\alpha|$ denotes the number of points contained in the cluster
$\alpha$.

Applying the Connolly Williams method to the expression (\ref{ce_excess})
rather than to (\ref{ce}) allows one to impose easily the  condition that
$\Delta f^s$ should be equal to zero for pure A and pure B. We thus
obtain the coefficients $f_{\alpha}$ only for clusters containing more
than one point, the coefficients $f_0$ and $f_1$ of the empty
and point clusters being then obtained by the relations
\begin{subequations}\begin{eqnarray}
f_0 &=& \frac{f_A+f_B}{2} - \sum_{\alpha, |\alpha|\ge2}
\frac{1+(-1)^{|\alpha|}}{2} D_{\alpha} f_{\alpha} ,
\\
f_1 &=& \frac{f_A-f_B}{2} - \sum_{\alpha, |\alpha|\ge2}
\frac{1-(-1)^{|\alpha|}}{2} D_{\alpha} f_{\alpha} .
\end{eqnarray}\end{subequations}

%The task of determining which clusters have to be used to obtain the best
%convergence is not an easy one.
%We have to try different sets, and then to check the convergence.
%This can be simplified by noting than clusters containing an odd or even
%number of points can be separated in the expansion. If we note by $\bar{s}$
%the symmetric of the structure $s$\footnote{For instance, if $s$ is A$_3$B
%in the structure L1$_2$, $\bar{s}$ will be B$_3$A in the structure
%L1$_2$.}, then only even
%clusters are present in the expansion of $f^s+f^{\bar{s}}$ and only odd
%clusters in the expansion of $f^s-f^{\bar{s}}$. This allows one to work on two
%smaller sets of clusters and thus facilitates the search of the best set of
%clusters.

%%%%%%%%%%%%%%%%%
\subsection{Relaxations}

The volume of a structure, like any other property, depends on its
configuration. But this volume enters directly in the cluster expansion as
the coefficients $f_{\alpha}$ have to be calculated for a given
volume.
As we are generally interested in the equilibrium properties, this leads to
some relaxations. In this study we will consider in two different ways these
volume relaxations, the globally and totally relaxed expansions
\cite{CAR87,SAN92}.

We first can make the cluster expansion explicitly volume dependent,
writing
\begin{equation}
f^s(V) = \sum_{\alpha} D_{\alpha} f_{\alpha}(V) \zeta_{\alpha}^s
\label{global_relaxed_ce} ,
\end{equation}
where the coefficients $f_{\alpha}(V)$ are obtained by calculating the
property $f^s$ for different structures at the same volume $V$, the other degrees of freedom
(shape of the unit cell and positions of atoms) being relaxed, and then by
using the Connolly-Williams method.
Doing such a cluster expansion for the energy, we can then deduce the
equilibrium volume associated with a given configuration by minimizing
with respect to the volume its energy as given by expression
(\ref{global_relaxed_ce}).
This is known as the globally relaxed scheme
and is based on the assumption that the volume occupied by every cluster is
independent on its configuration. Such an assumption is questionable in
cases where there is a significant difference between the atomic volumes of
the constituent elements as in the Al-Zr system.

Another way to consider relaxations of the volume is to calculate the
coefficients $f_{\alpha}$ from the equilibrium values $f^s(V_0^s)$
 where everything included the volume is allowed to relax.
The coefficients $f_{\alpha}$ are
then volume independent and the values predicted by the expansion are
directly the ones at equilibrium.
Such a treatment is called a totally relaxed expansion. This
expansion is still rigorous from a mathematical point of view since the relaxations
are themselves functions of the configuration, so the relaxed
structures will also be.

%%%%%%%%%%%%%%%%%%%%
\subsection{Results}
\subsubsection{Total relaxations}

\begin{figure}[!hbt]
\begin{tabular}{cccc}

\subfigure[\{3,1\}]{
\includegraphics[width=0.1\textwidth]{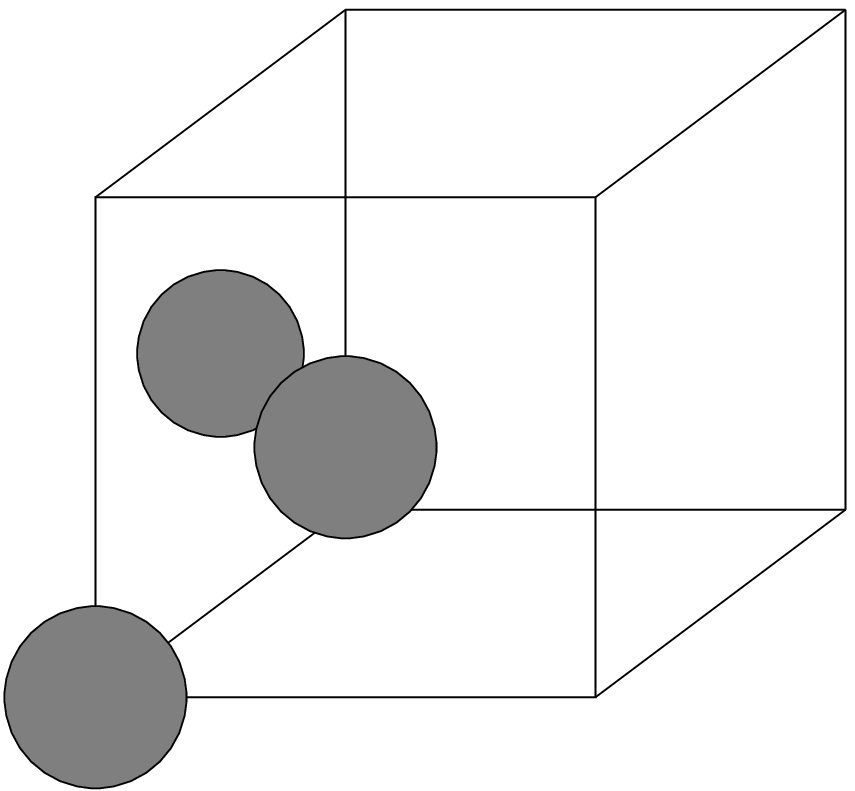}} &
\subfigure[\{3,2\}]{
\includegraphics[width=0.1\textwidth]{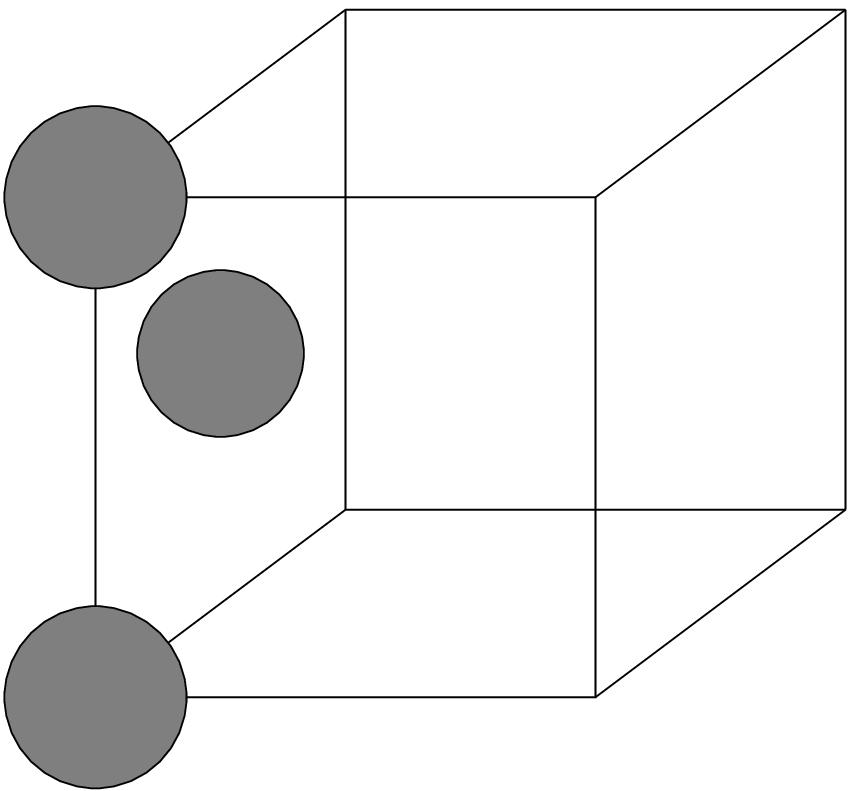}} &
\subfigure[\{3,3\}]{
\includegraphics[width=0.1\textwidth]{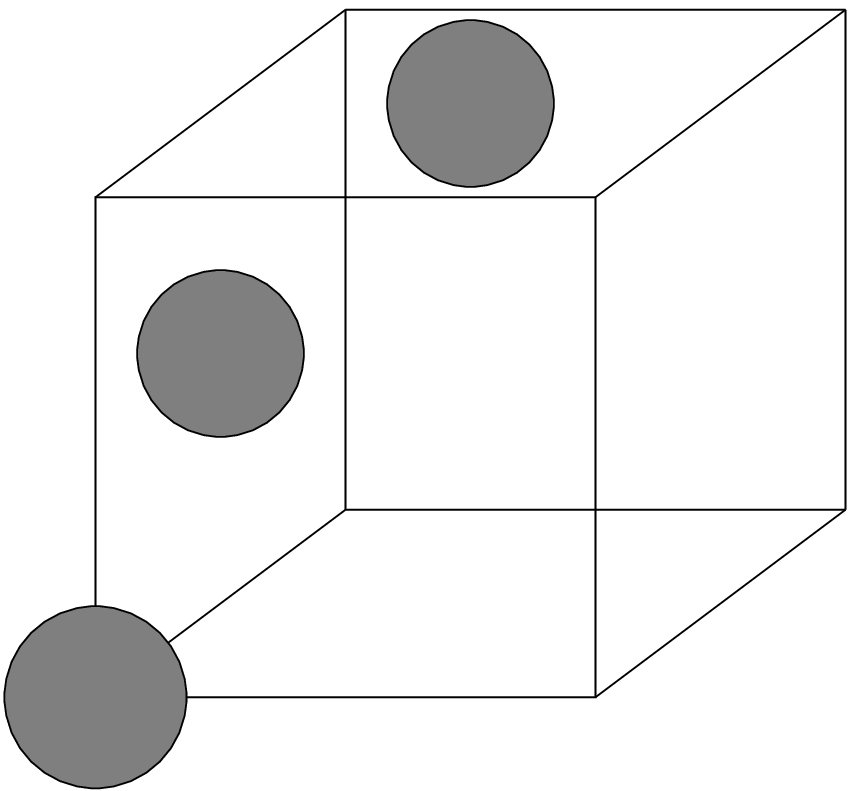}} &
\subfigure[\{3,4\}]{
\includegraphics[width=0.1\textwidth]{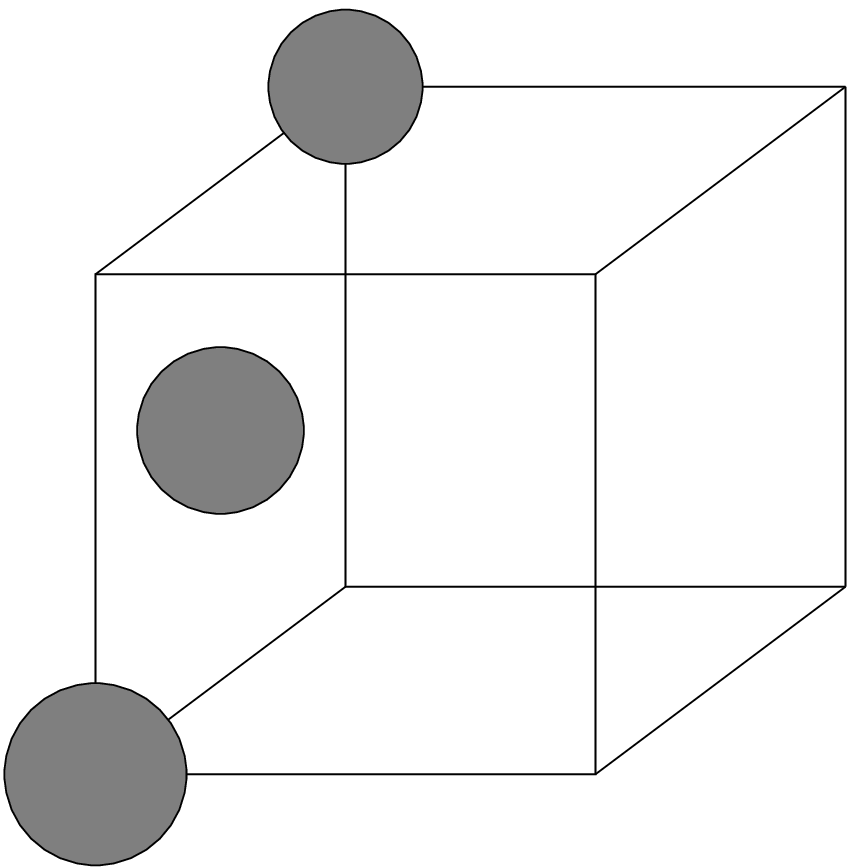}}\\
\subfigure[\{3,5\}]{
\includegraphics[width=0.1\textwidth]{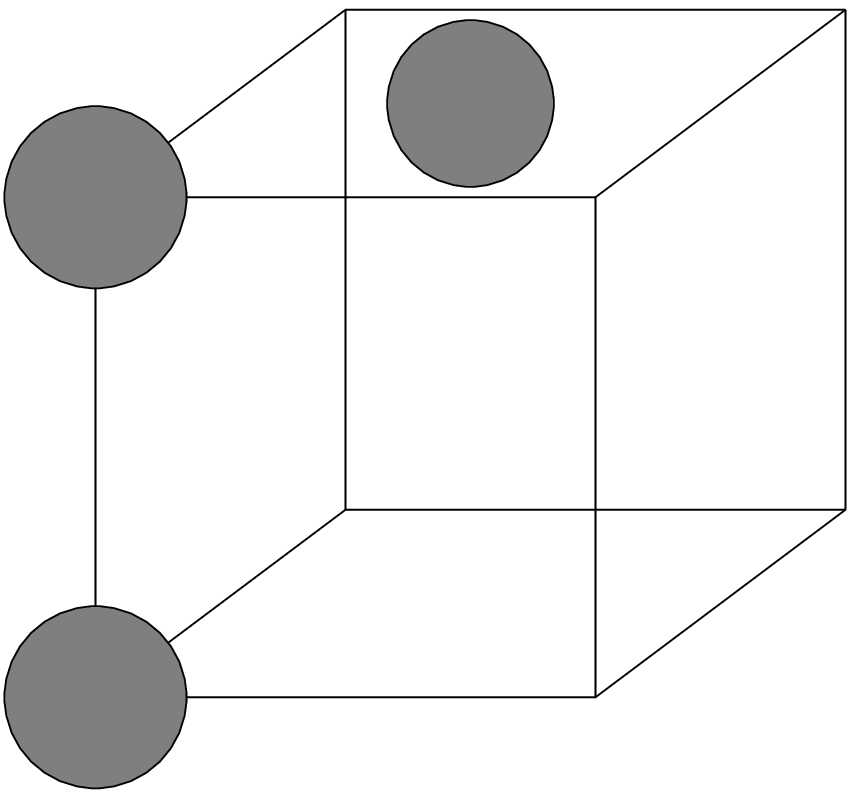}} &
\subfigure[\{3,6\}]{
\includegraphics[width=0.1\textwidth]{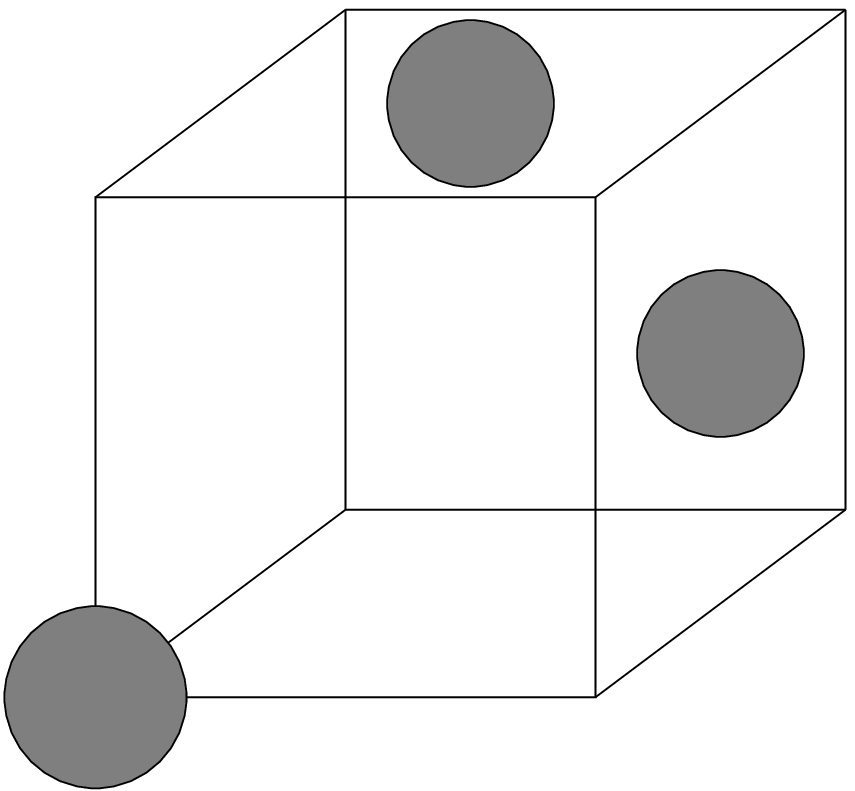}} &
\subfigure[\{3,7\}]{
\includegraphics[width=0.1\textwidth]{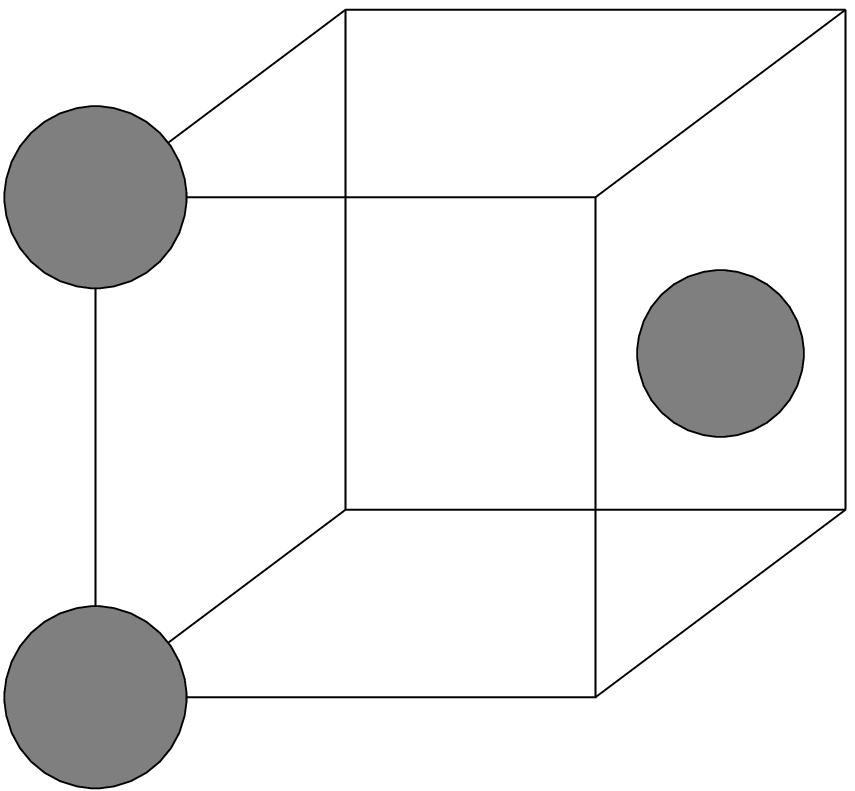}} &
\subfigure[\{4,1\}]{
\includegraphics[width=0.1\textwidth]{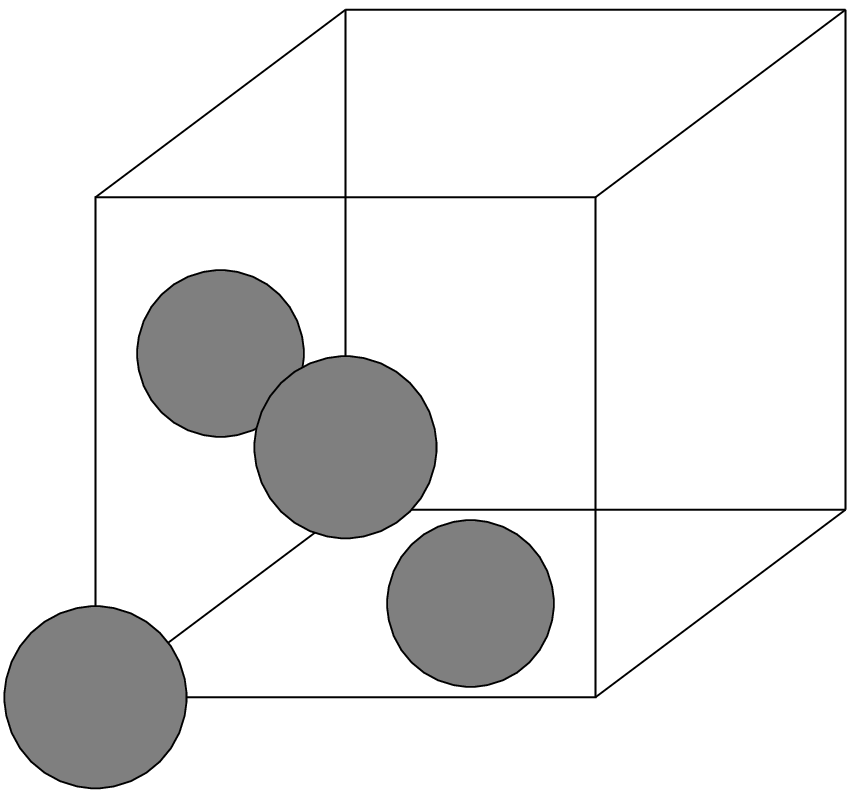}}

\end{tabular}
\caption{Definition of the three and four points clusters on an fcc lattice used for the expansion.}
\label{clusters_fig}
\end{figure}

We first made a cluster expansion of the equilibrium volume, the bulk
modulus, and the formation energy for the Al-Zr system on an
fcc lattice: we are thus doing three different totally relaxed expansions.
To perform the least square fit of the expansion, we used the 26
structures for which these equilibrium properties were obtained from our
FP-LMTO calculations (Table \ref{AlZr_relaxed_lda}).
The best agreement between our \abinitio calculations and their expansion
has been obtained when using 17 clusters:
the empty cluster \{0\}, the point cluster \{1\},
the pairs from first to seventh nearest neighbours \{2,1\}\ldots\{2,7\},
seven triangles \{3,1\}\ldots\{3,7\} presented in figures \ref{clusters_fig}~a-g
and the tetrahedron of first nearest neighbours \{4,1\} (Fig. \ref{clusters_fig}~h).
The values of the coefficients obtained for these three totally relaxed
expansions are presented in table \ref{coef_ce_total} and the errors compared
to the direct calculation for the 25 structures in table \ref{error_ce_total}.

\begin{table}[!hbt]

\caption[Cluster expansion of the equilibrium volume, bulk modulus, and
formation energy at equilibrium in the total relaxations scheme]
{Cluster expansion of the equilibrium volume (coefficients $V_\alpha$), bulk
modulus ($B_\alpha$), and formation energy ($J_\alpha$) in
the total relaxation scheme.}
\label{coef_ce_total}

\begin{ruledtabular}
\begin{tabular}{c.{-1}.{-1}.{-1}.{-1}}
&&\multicolumn{1}{c}{$V_\alpha$}        & \multicolumn{1}{c}{$B_\alpha$}&
\multicolumn{1}{c}{$J_\alpha$}  \\
\raisebox{1.5ex}[0pt]{Cluster}  &
\multicolumn{1}{c}{\raisebox{1.5ex}[0pt]{$D_{\alpha}$}} &
\multicolumn{1}{c}{(\AA$^3$/atom)}      & \multicolumn{1}{c}{(GPa)}     &
\multicolumn{1}{c}{(mRy/atom)}  \\
\hline
\{0\}   & 1     & 18.587 &  98.15 & -625.05 \\
\{1\}   & 1     & -3.230 & -11.12 &  419.87 \\
\{2,1\} & 6     &  0.149 &  -1.12 &    3.69 \\
\{2,2\} & 3     & -0.128 &   1.88 &   -3.86 \\
\{2,3\} & 12    & -0.013 &  -0.09 &    0.07 \\
\{2,4\} & 6     & -0.027 &  -0.11 &   -0.15 \\
\{2,5\} & 12    & -0.037 &   0.07 &    0.16 \\
\{2,6\} & 4     &  0.009 &  -0.30 &    0.93 \\
\{2,7\} & 24    &  0.014 &  -0.17 &    0.18 \\
\{3,1\} & 8     &  0.013 &  -0.14 &    1.74 \\
\{3,2\} & 12    & -0.031 &  -0.40 &   -0.45 \\
\{3,3\} & 24    & -0.001 &  -0.14 &   -0.55 \\
\{3,4\} & 6     &  0.023 &   0.21 &   -0.31 \\
\{3,5\} & 24    &  0.004 &   0.21 &   -0.33 \\
\{3,6\} & 24    &  0.010 &   0.14 &    0.22 \\
\{3,7\} & 24    &  0.005 &   0.06 &    0.54 \\
\{4,1\} & 2     &  0.030 &  -0.55 &    0.88 \\
\end{tabular}
\end{ruledtabular}

\end{table}

\begin{table}[!hbt]

\caption{Deviations for the cluster expansion of the equilibrium volume
($\delta V_0$), the bulk modulus ($\delta B$), and the formation energy
($\delta E^{form}$) in the total relaxations scheme.}
\label{error_ce_total}

\begin{ruledtabular}
\begin{tabular}{l.{-1}.{-1}.{-1}}

&\multicolumn{1}{c}{$\delta V_0$}       &
\multicolumn{1}{c}{$\delta B$}          &
\multicolumn{1}{c}{$\delta E^{form}$}   \\
\raisebox{1.5ex}[0pt]{} &
\multicolumn{1}{c}{(\AA$^3$/atom)}      &
\multicolumn{1}{c}{(GPa)}               &
\multicolumn{1}{c}{(mRy/atom)}          \\
\hline
Al (fcc)                              &  0.      &  0.     &  0.    \\
Al$_{31}$Zr                       &  0.052 &  0.30  &  0.42 \\
Al$_{15}$Zr                       &  0.035 &  0.83  & -1.02  \\
Al$_8$Zr (NbNi$_8$)          &  0.157 & -1.05  &  4.01  \\
Al$_7$Zr (D1)                    &  0.018 &  0.31   &  0.87  \\
Al$_4$Zr (D1$_{\textrm{a}}$)& -0.012 & -1.62  &  1.96  \\
Al$_3$Zr (L1$_2$)              & -0.109 & -0.02  & -0.21  \\
Al$_3$Zr (DO$_{22}$)         & -0.029 &  0.63  & -0.58  \\
Al$_3$Zr (DO$_{23}$)         & -0.079 &  0.73  & -2.10  \\
Al$_2$Zr ($\alpha$)            & -0.017 &  0.10  & -0.36  \\
Al$_2$Zr ($\beta$)              & -0.028 &  0.52  & -1.11  \\
Al$_2$Zr ($\gamma$)          & -0.041 &  0.60  & -1.38  \\
AlZr (L1$_0$)                     &  0.219 & -1.18  &  1.96  \\
AlZr (L1$_1$)                     &  0.323 & -0.86  &  0.60  \\
AlZr (CH40)                       &  0.077 & -0.11  &  0.56  \\
AlZr (D4)                           & -0.348 &  0.64  & -1.20  \\
AlZr (Z2)                           &  0.       &  0.41  &  0.49  \\
Zr$_2$Al ($\alpha$)            &  0.011 &  0.03  &  0.16  \\
Zr$_2$Al ($\beta$)              & -0.007 &  0.46  & -0.71  \\
Zr$_2$Al ($\gamma$)          &  0.002 &  0.49  & -0.59  \\
Zr$_3$Al (L1$_2$)              & -0.095 &  1.21  & -0.32  \\
Zr$_3$Al (DO$_{22}$)         & -0.038 &  2.01  & -1.29  \\
Zr$_3$Al (DO$_{23}$)         & -0.061 & -2.03  & -0.69  \\
Zr$_4$Al (D1$_{\textrm{a}}$) & -0.012 & -1.62  &  1.96  \\
Zr$_7$Al (D1)                     &  0.085  &  0.32  &  1.78  \\
Zr (fcc)                              &   0.      &  0.      &  0.    \\
&&&\\
Standard deviation              &  0.116 &  0.91  &  1.33  \\

\end{tabular}
\end{ruledtabular}
\end{table}

Looking at the cluster expansion of the formation  energy,
it can be seen that the maximum difference
between the energy given by the expansion and the one directly obtained
from the FP-LMTO calculations is 4.0~mRy/atom and that the standard deviation
is 1.4~mRy/atom. We did not manage to find a better set of clusters producing
a smaller error:
as we still had more structures to fit than clusters, we tried to include more clusters
like the pair to the eighth nearest neighbour but this did not improve the difference
between our FP-LMTO calculations and their expansion.
Such an error does not allow one to reproduce the relative stability
between different ordered compounds at a same concentration,
for instance between the phases L1$_2$, DO$_{22}$, and DO$_{23}$ of Al$_3$Zr.
But as we are interested in using the cluster expansion only for the solid solution Al-Zr,
this is not a problem: for perfectly ordered compounds we can directly use
the results of our ab-initio calculations.

For the equilibrium volume, we can compare the accuracy of the cluster
expansion with the one of the Vegard's law which assumes a
linear relation between the atomic volume and the concentration.
The standard deviation of the Vegard's law is 0.427~\AA/atom.
For none of the considered structures such an important error
occurs, and we have thus obtained an important improvement by not considering
only the empty and point cluster as one does in the Vegard's law.

For the bulk modulus, the accuracy of our FP-LMTO being of the order of
1~GPa, here too we can consider the convergence of the cluster expansion to
be good.

%%%%%%%%%%%%%%%%%%
\subsubsection{Global relaxations}

Using the same sets of clusters and structures, we expanded
the ground state energy in 21 different volumes between 14 and 24
\AA$^3$/atom. For each structure, these 21 expansions gave the ground state
energy of the relaxed structures at the corresponding fixed volume,
and we then used these results to obtain
the volume, the bulk modulus, and the ground state energy at
equilibrium by fitting the Rose equation of state \cite{ROS84}.

\begin{table}[!hbp]

\caption{Deviations for the cluster expansion of the equilibrium volume ($\delta
V_0$), the bulk modulus ($\delta B$), and the formation energy
($\delta E^{form}$) obtained in the global relaxations scheme.}
\label{error_ce_global}

\begin{ruledtabular}
\begin{tabular}{l.{-1}.{-1}.{-1}}

&\multicolumn{1}{c}{$\delta V_0$}       &
\multicolumn{1}{c}{$\delta B$}          &
\multicolumn{1}{c}{$\delta E^{form}$}   \\
\raisebox{1.5ex}[0pt]{} &
\multicolumn{1}{c}{(\AA$^3$/atom)}      &
\multicolumn{1}{c}{(GPa)}               &
\multicolumn{1}{c}{(mRy/atom)}          \\
\hline
Maximal deviation                   & -0.351 & -2.26 &  5.15\\
Standard deviation              &  0.120&  0.94 &  1.44 \\

\end{tabular}
\end{ruledtabular}
\end{table}

The maximal and standard deviations between the properties deduced from the
expansion and the ones directly obtained from the FP-LMTO calculations are
shown in table \ref{error_ce_global}.
They are close to what we previously obtained in the total relaxation scheme.
Actually, we did not get any important difference between the
results obtained according to the way volume relaxations are considered.
For each structure the deviation is quite the same in both cases,
this being true for the formation  energy
as well as for the equilibrium volume and the bulk modulus.
Such a result could not have been easily predicted as the size difference between
Al and Zr is quite important: the atomic volumes given by our calculations
for the fcc structures of Al and Zr are respectively 15.82 and 21.70~\AA${^3}$
(Table \ref{AlZr_relaxed_lda}).
This proves that the globally and locally relaxed expansions are equivalent.

As the totally relaxed expansion only gives us one set of coefficients for
the whole range of volumes, it is more convenient and we will use this
expansion in the following.

%%%%%%%%%%%%%%%%%%%%%%%%%%%%%%%%%%%%%%%%%%%

\section{Finite temperature properties}

At finite temperature, the vibrational  and electronic contributions
as well as the configurational entropy have
to be included in the description of the system.
Considering two different time scales, a slow one for the configurational
effects and a much faster for vibrations and electronic excitations
\cite{DEF94}, we define a vibrational and electronic free energy,
$F^{vib}(\SIGMA)$ and $F^{el}(\SIGMA)$, both depending on the
configuration.
Using the variational principle, the free energy is obtained by minimizing
the functional
\begin{equation}
F[\rho] = \langle E_0 \rangle
+ \langle F^{vib} \rangle
+ \langle F^{el} \rangle
+ k_B T \langle \ln\rho \rangle
\label{functionalF} ,
\end{equation}
where $k_B$ is the Boltzmann constant and $\rho$ the density matrix.

The cluster expansion of the formation energy at $T=0$~K gives us an
expression for the cohesive part of the functional of Eq. (\ref{functionalF}).
We do not have to take into account any variation of the lattice parameter with temperature
as we chose to work in the harmonic approximation:
Ozoli\c{n}\v{s} and Asta\cite{OZO01} showed on the solubility limit of Sc in Al that there was only a small
improvement when going from the harmonic to the quasiharmonic approximation.
Similar expressions have to be found for the electronic and vibrational
parts of the expression (\ref{functionalF}).
The minimization of $F[\rho]$ with respect to $\rho$
will then be done in the Bragg-Williams approximation.

%%%%%%%%%%%%%%%%%%%%%%%%%%%%%%%%%%%
\subsection{Electronic free energy}

At a temperature of 0~K, all electronic states of energy below the Fermi
level $\epsilon_f$ are occupied, whereas the ones above are empty.
At finite temperature, the electrons close to the Fermi levels can be
promoted to states of higher energies according to the Fermi-Dirac
distribution $f(\epsilon,T)$.
The electronic excitations induce a change of the charge density and thus
of the effective potential of the one electron Hamiltonian. This leads the
electronic density of states (DOS) $n(\epsilon)$ to be temperature
dependent.
But the changes induced on the total energy
and on the entropy by this temperature dependence of the electronic DOS
are small \cite{WOL95}.
We thus assumed the electronic DOS to be
temperature independent and equal to the one obtained at $T=0$.
The energy change $\Delta E^{el}(T)$ and the entropy $S^{el}(T)$ due to
electronic excitations are then
\begin{subequations}
\begin{eqnarray}
\Delta E^{el} &=& \int_{-\infty}^{\infty}\epsilon n(\epsilon)
\left[ f(\epsilon,T) - f(\epsilon,0) \right] \deriv\epsilon ,  \\
%\end{equation}
%\begin{eqnarray}
S^{el}  &=& -k_B\int_{-\infty}^{\infty}n(\epsilon)\left\{
 f(\epsilon,T)\ln[f(\epsilon,T)] \right. \nonumber\\
 & & \left. +[1-f(\epsilon ,T)]\ln[1-f(\epsilon,T)]
\right\}\deriv\epsilon .
\end{eqnarray}
\end{subequations}

  % Electronic free energy

\begin{figure}[!hbt]
\begin{center}
\includegraphics[width=0.4\textwidth]{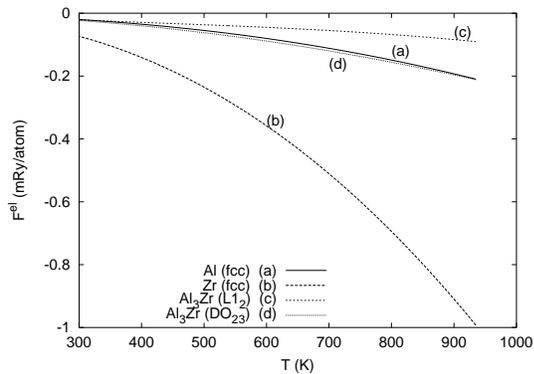}
\caption{Electronic free energy, $F^{el}=\Delta E^{el}-TS^{el}$.}
\label{plot_Fel}
\end{center}
\end{figure}

We calculated the electronic contribution to the free energy,
$F^{el}= \Delta E^{el} - T S^{el}$, for the
structures Al (fcc), Zr (fcc), Al$_3$Zr (L1$_2$), and  Al$_3$Zr (DO$_{23}$)
(Fig. \ref{plot_Fel}).
In the range of temperature of interest, \ie below 1000~K, this electronic
contribution is smaller than 1 mRy/atom, and so is the excess free energy
associated.
This is the same range of order as the accuracy of the cluster expansion
of the formation energy.
We thus chose to neglect this contribution to the free energy.

%\clearpage
%%%%%%%%%%%%%%%%%%%%%%%%%%%%%%%%%%%
\subsection{Vibrational free energy}

We studied the vibrational effects in the harmonic approximation, comparing
the ability of the Debye model with a phonon calculation to predict the
thermodynamic properties.

%%%%%%%%%%%%%%%%
\subsubsection{Phonon calculation}

A calculation of the phonon DOS $n(\omega)$ allows one to compute the
vibrational free energy. For temperatures higher than 300~K, it is enough
to consider only its high temperature expression
\begin{eqnarray}
F^{vib} &=& k_B T \left[ -3 \ln(k_BT)
+ \int_0^\infty \ln(\hbar\omega)n(\omega) \deriv\omega \right]
\nonumber \\&&
%+ \left. \frac{1}{24(k_B T)^2} \int_0^{\infty} (\hbar\omega)^2n(\omega)
+ O \left(\frac{1}{T}\right) .
\end{eqnarray}

Phonon DOS were calculated for Al~(fcc), Zr~(fcc), Al$_3$Zr~(L1$_2$), and
Al$_3$Zr~(DO$_{23}$) in the linear response theory framework \cite{BAR87}.
We used energy independent MTO as a basis for representing the first order
correction to the one electron wave functions in the implementation developed
by Savrasov \cite{SAV92b,SAV96}.
These calculations were performed in the LDA using the parametrization of
Moruzzi-Janak-Williams \cite{MOR78}.
The radii of the muffin-tin spheres were taken equal to the ones of the
band structure calculation.
For valence states, the basis used was the same whereas the 4s and 4p states
of Zr were treated in two different panels with the respective kinetic
energies $\kappa^2$ 2.7 and 1.1 mRy.
For fcc structures, phonon frequencies were calculated on a grid of
$8\times8\times8$ wave vectors $\vec{q}$ leading to 29 points in the
irreducible Brillouin zone (IBZ), for L1$_2$ a grid of $5\times5\times5$
leading to 10 points in the IBZ was used, and for DO$_{23}$ a grid of
$4\times4\times4$ wave vectors leading to 13 points.

\begin{figure}[!hbtp]
\includegraphics[width=0.45\textwidth]{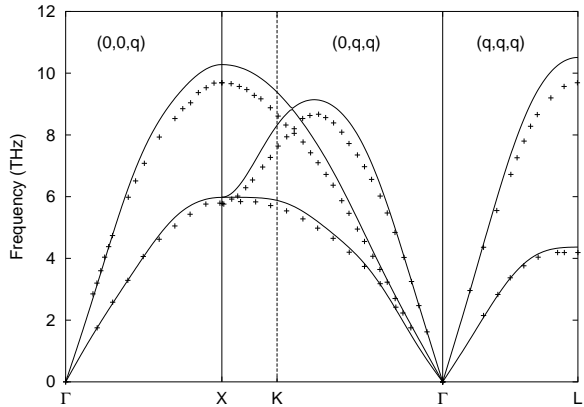}
\caption{Calculated phonons dispersion for Al~fcc in solid line compared to experimental data
(Ref.~\cite{STE66,LANDOLT2}) denoted by crosses.}
\label{Al_phonon_dispersion}
\end{figure}

The calculated phonons dispersion for Al~fcc is compared in figure \ref{Al_phonon_dispersion}  to the
measurements of Stedmam \etal \cite{STE66,LANDOLT2} for three different high-symmetry directions.
We see that our calculation overestimates phonon frequency.
Other phonon calculations for Al~fcc\cite{QUO92,GIR95,BAU98} using the linear response
theory too obtained a better agreement with experimental data.
They all used a plane waves basis in the pseudopotential framework, but the use of an energy
independent MTO as a basis does not seem to be the reason of the discrepancy with experimental data
in our case, as Savrasov showed for Nb\cite{SAV92b}as well as for NbC and Si\cite{SAV96}
that this basis was well-suited to obtain phonon dispersion.

\begin{figure*}[t]
\subfigure[Al (fcc): (a) from ref. \cite{COW74,LANDOLT2}
and (b) from ref. \cite{GIL66,LANDOLT2}]{
\includegraphics[width=0.4\linewidth]{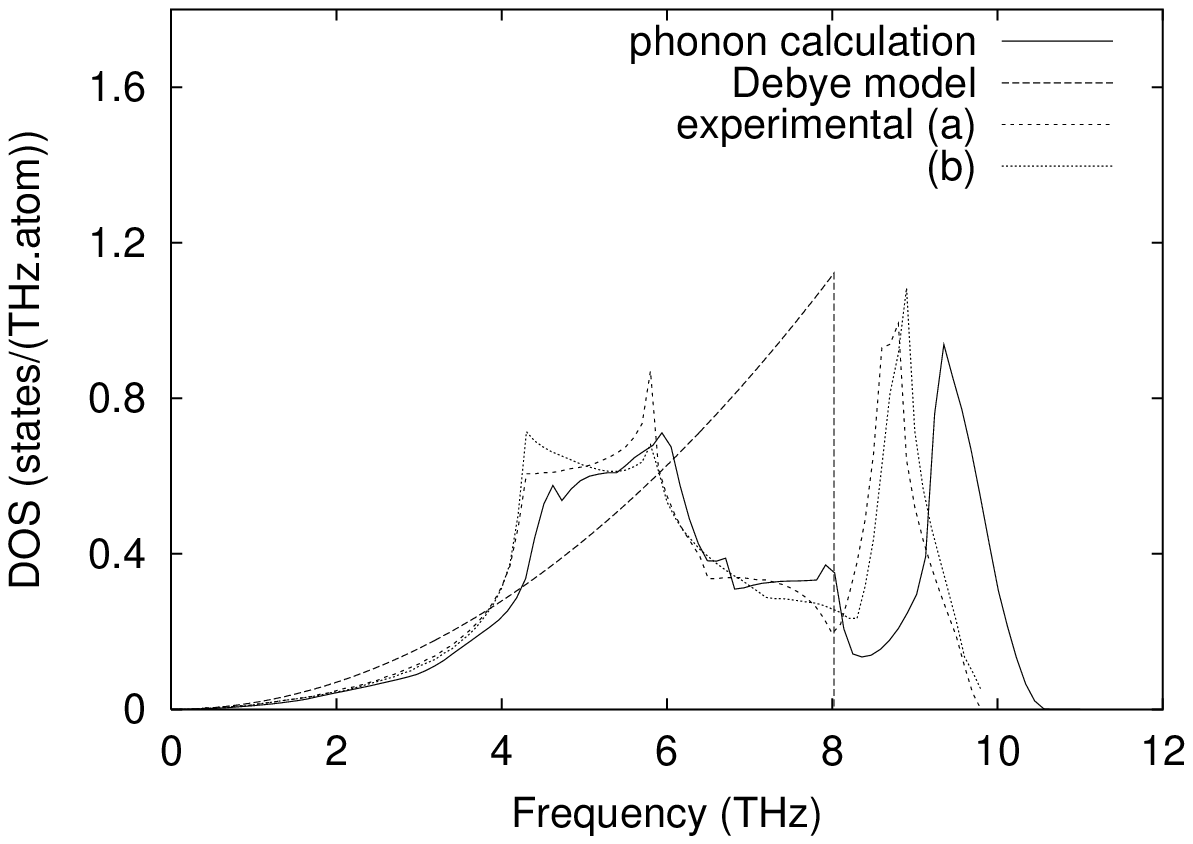}
}
\subfigure[Zr (fcc)]{
\includegraphics[width=0.4\linewidth]{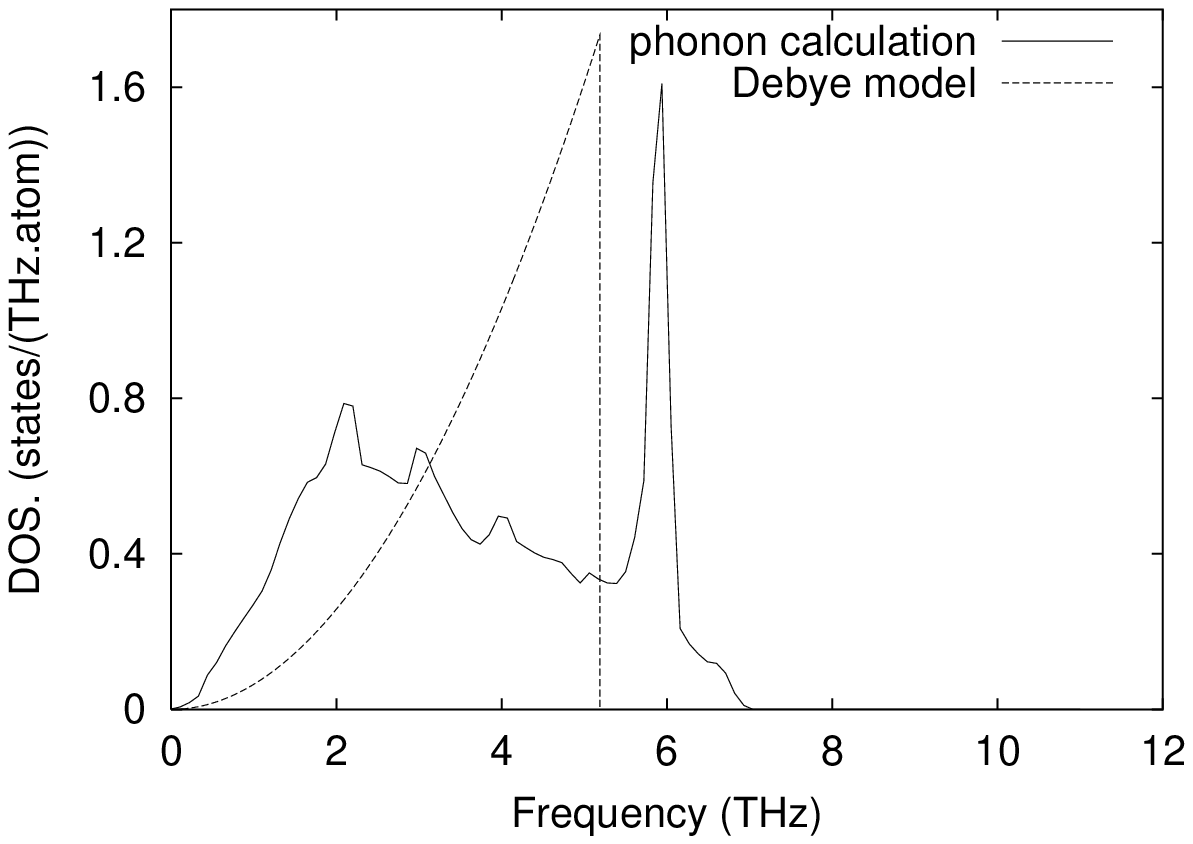}
}
\subfigure[Al$_3$Zr (L1$_2$)]{
\includegraphics[width=0.4\linewidth]{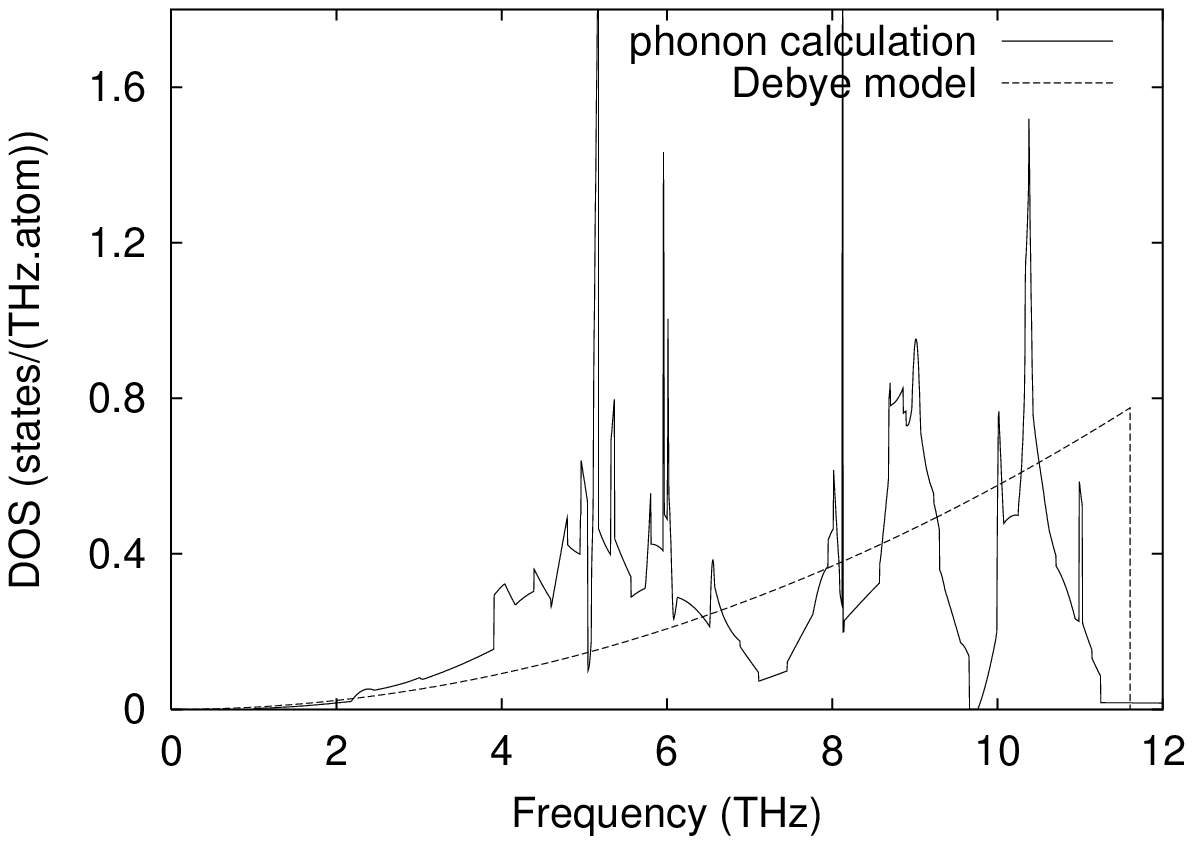}
}
\subfigure[Al$_3$Zr (DO$_{23}$)]{
\includegraphics[width=0.4\linewidth]{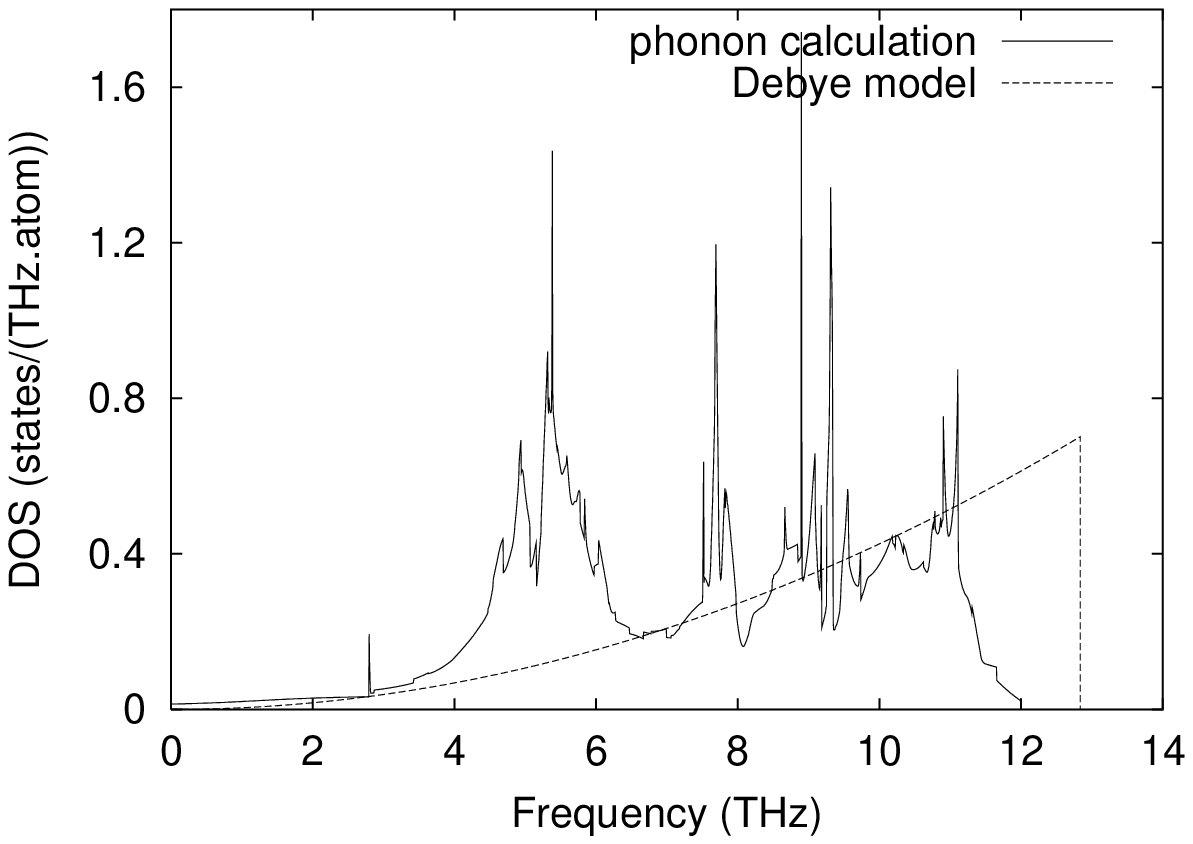}
}
\caption{Phonon density of state.}
\label{phonon_dos}
\end{figure*}

The phonon DOS obtained from these calculations for Al~(fcc), Zr~(fcc),
Al$_3$Zr~(L1$_2$), and Al$_3$Zr~(DO$_{23}$)
are presented in Fig. \ref{phonon_dos}.
For Al~(fcc), we compared our calculated phonon DOS with experimental ones.
Experimental DOS \cite{COW74,GIL66,LANDOLT2} were obtained by means
of a Born-von Karman model. Force constants were fitted up to the eighth
nearest neighbours in order to reproduce the phonon measurements in high-symmetry
directions of Stedmam \etal \cite{STE66,LANDOLT2}, the Born-von Karman
model being used then to compute the frequency distribution.
We can see on the phonon DOS too that our calculated frequencies are slightly too high.
Nevertheless, the shape of the frequency distribution is correct.

%%%%%%%%%%%%%%%
\subsubsection{Debye model}

The Debye model assumes a linear dispersion between the phonon frequency
and its wave vector. This leads to the following high temperature expression
of the vibrational free energy
\begin{equation}
F^{vib} = k_BT\left[ -1 +3 \ln \left(\frac{\theta_D}{T}\right)
%+ \frac{3}{40} \left(\frac{\theta_D}{T}\right)^2
\right]
+ O \left(\frac{1}{T}\right)
\label{Debye_thermo_highT} ,
\end{equation}
where the Debye temperature $\theta_D$ is obtained from the elastic
constants of the structures \cite{ALE65}.

The elastic constants were obtained by means of FP-LMTO calculations using the
same set of parameters as for the formation energy calculations.
The unit cell of the crystal was deformed around its
equilibrium position in order to obtain the second derivative of the energy
at its minimum which can be then related to the elastic tensor
\cite{SOD93,FAS95}.
During this deformation, no relaxation was allowed.
For the DO$_{23}$ structure, the $c/a$ ratio and the position $\delta_{\textrm{Al}}$
and $\delta_{\textrm{Zr}}$ of the atoms were frozen at their equilibrium value.
For some of the deformations, we checked that these relaxations did not
change much the values of elastic constants.
Moreover, as we are lowering the symmetry of the structure by deforming it,
some new degrees of freedom can appear,
but we did not consider either these ones.
% For all the structures we already knew the bulk modulus, so only two more
% relations are needed for structures having cubic symmetry, four for
% hexagonal, and five for tetragonal.

The elastic constants calculated with the FP-LMTO are
compared to the experimental ones in table \ref{elasticity1}.
The discrepancy between the calculated and experimental
constants is in the order of 10\%.
This leads to some differences between the Debye temperatures
obtained from these calculated constants and the ones obtained
from the experimental constants, but the relative positions of these
temperatures are correctly predicted.
% We can compare the Debye temperature obtained from the elastic constants to
% the ones directly obtained by a calorimetric measurement of the specific
% heat:
% for Al $\theta_D=428$~K \cite{PHI59} and for Zr $\theta_D=310$~K
% \cite{WOL57}. These data are really close to what we obtained from the
% experimental elastic constants, showing that for pure metals a calculation from the elastic
% constants is a good way to obtain the Debye temperature, the precision of
% the result being directly related to the precision of the calculation of
% the elastic constants.

\begin{table*}[!hbtp]

\caption[Elastic constants $C_{ij}$ calculated with the FP-LMTO compared
to experimental values for Al (fcc), Al$_3$Zr (DO23), and Zr (hcp), and
Debye temperature associated]
{Elastic constant $C_{ij}$ (in GPa) calculated with the FP-LMTO compared
to experimental values for Al (fcc), Al$_3$Zr (DO23), and Zr (hcp), and
Debye temperature $\theta_D$ associated.
Debye temperatures obtained by calorimetric measurements of the specific heat,
when available, are given in brackets.}
\label{elasticity1}

\begin{ruledtabular}
\begin{tabular}{ll.{-1}.{-1}.{-1}.{-1}.{-1}.{-1}cc}

&&\multicolumn{1}{c}{$C_{11}$}
&\multicolumn{1}{c}{$C_{33}$}
&\multicolumn{1}{c}{$C_{12}$}
&\multicolumn{1}{c}{$C_{13}$}
&\multicolumn{1}{c}{$C_{44}$}
&\multicolumn{1}{c}{$C_{66}$}
&\multicolumn{2}{c}{$\theta_D$ (K)}\\
\hline
Al (fcc)        & FP-LMTO &
101.5 & \cdots& 70.4  & \cdots& 31.7  & \cdots & 385\\
 &exp.\cite{LANDOLT,KAM64}
        & 114.3 & \cdots& 61.9  & \cdots& 31.6  & \cdots & 431 & (428) \cite{PHI59} \\
\\
Al$_3$Zr (DO23) & FP-LMTO &
 215.3 & 228.2 & 54.1  & 33.3  & 103.2 & 123.5 & 616\\
& exp.\cite{NAK91}\footnote{measured at room temperature}
        & 208.8 & 208.3 & 70.5  & 49.1  & 87.2  & 102.2 & 575\\
\\
Zr (hcp) & FP-LMTO &
153.1 & 171.2 & 63.4  & 76.5  & 22.4  & 44.9  & 262\\
& exp.\cite{LANDOLT,FIS64}
        & 155.4 & 172.5 & 67.2  & 64.6  & 36.3  & 44.1  & 299 & (310) \cite{WOL57}\\

\end{tabular}
\end{ruledtabular}
\end{table*}

In table \ref{elasticity2}, we show Debye temperatures obtained from a
calculation of the elastic tensor for cubic structures fcc, D1, and L1$_2$
of the Al-Zr system. The structure D4 of AlZr is cubic too, but this
phase was found to be mechanically unstable through a Bain deformation
path and cannot be used to calculate a Debye temperature.

\begin{table}[!hbt]

\caption[elastic constants for Al-Zr compounds of cubic symmetry
and Debye temperature associated]
{Elastic constants $C_{ij}$ (in GPa) for Al-Zr compounds of cubic symmetry
and Debye temperature $\theta_D$ associated.}
\label{elasticity2}

\begin{ruledtabular}
\begin{tabular}{l.{-1}.{-1}.{-1}c}

& C_{11} & C_{12} & C_{44} & $\theta_D$ (K) \\
\hline
Al (fcc)                & 101.5 & 70.4  & 31.7  & 385   \\
Al$_7$Zr (D1)           & 136.5 & 62.7  & 45.8  & 449   \\
Al$_3$Zr (L1$_2$)       & 187.3 & 55.7  & 95.1  & 557   \\
Zr$_3$Al (L1$_2$)       & 163.8 & 79.3  & 86.5  & 388   \\
Zr$_7$Al (D1)           & 136.3 & 84.4  & 56.6  & 300   \\
Zr (fcc)                & 121.4 & 87.1  & 45.7  & 249   \\

\end{tabular}
\end{ruledtabular}
\end{table}

%%%%%%%%%%%%%%%%%%%%%%%%%%
\subsubsection{Comparison for ordered compounds}

As we calculated the phonon spectrum for Al$_3$Zr for the stable structure
DO$_{23}$ and the metastable one L1$_2$, we were able to compare the
excess vibrational free energy $\Delta F^{vib}$ obtained from the phonon
DOS and the Debye model, the reference phases being Al~(fcc) and Zr~(fcc)
 (high temperature expressions are given in table \ref{highT_table}).
We thus see that the Debye model makes an important error in predicting
this excess free energy as it overestimates it by a factor $\sim2$.
This error comes from the inability of the Debye model to reproduce the phonon DOS as shown by
figure \ref{phonon_dos}.
Moreover the phonon calculation shows that the two considered structures
of Al$_3$Zr should have the same vibrational free energy which is not
correctly predicted by the Debye model.
This error of the Debye model would lead to a stabilization of the phase L1$_2$
at high temperatures ($T\gtrsim$905~K) which is not true experimentally.
In order to correctly describe the relative stability of these two ordered phases
of Al$_3$Zr we cannot use the Debye model, but we need the phonon calculation.

\begin{table}[!hbtp]

\caption{Comparison of the high temperature expressions of the vibrational
free energy obtained with the phonon calculation and the Debye model.}
\label{highT_table}

\begin{ruledtabular}
\begin{tabular}{llccl}

Al$_3$Zr (L1$_2$) & Phonons & $\Delta F^{vib}$&=&
$0.85 k_B T   + O\left(\frac{1}{T}\right)$ \\
%$k_BT\left[0.85 +\frac{6088}{T^2} + O\left(\frac{1}{T^4}\right)\right]$\\
&Debye  &       &=&  $1.44 k_B T + O\left(\frac{1}{T}\right)$ \\
%$k_BT\left[1.44 +\frac{13767}{T^2} + O\left(\frac{1}{T^4}\right)\right]$\\
\hline
Al$_3$Zr (DO$_{23}$) & Phonons & $\Delta F^{vib}$&=&
$0.85 k_B T   + O\left(\frac{1}{T}\right)$ \\
%$k_BT\left[0.85 +\frac{6655}{T^2} + O\left(\frac{1}{T^4}\right)\right]$\\
&Debye  &       &=&  $1.74 k_B T + O\left(\frac{1}{T}\right)$ \\
%$k_BT\left[1.74 +\frac{18956}{T^2} + O\left(\frac{1}{T^4}\right)\right]$\\

\end{tabular}
\end{ruledtabular}

\end{table}

%%%%%%%%%%%%%%%%%%%%%%%%%%
\subsubsection{Cluster expansion for the disordered phase}

For the vibrational free energy of the disordered phase, we made a cluster expansion of
the vibrational free energies of several ordered structures.
As the Debye model only requires the calculation of the elastic tensor, which
is much more faster than a calculation of the whole phonon spectrum, we used
it to calculate the vibrational free energy of these ordered compounds
(Debye temperatures in Table \ref{elasticity2}).
By doing so we saw previously that we overestimates $\Delta F^{vib}$, but
a calculation of the phonon spectrum is not conceivable for
a number of structures large enough to fit the cluster expansion.
We have then to accept such an error.

Looking at the high temperature expression of the vibrational free energy given by the Debye model
(Eq. \ref{Debye_thermo_highT}),
we can make the following cluster expansion
\begin{equation}
3\ln \theta_D -1 = \sum_\alpha D_{\alpha} J_{\alpha} \zeta_{\alpha} ,
\end{equation}
which allows us to write the vibrational free energy as
\begin{equation}
F^{vib}= k_B T \left[ \sum_\alpha D_{\alpha} J_{\alpha} \zeta_{\alpha}
-3 \ln T \right]
\label{Fvib_ce} .
\end{equation}
By doing so, the temperature dependence of the free energy is really simple
and we do not have to make a cluster expansion of the free energy
at every temperature.

We only used four clusters in the truncated expansion: the empty cluster
\{0\}, the point cluster \{1\}, the pair \{2,1\} of first nearest
neighbours, and the triangle \{3,1\} of first nearest neighbours.
The eight structures of table \ref{elasticity2} were used to fit the
coefficients of the expansion.
The results of this expansion are presented in table \ref{ce_vib}(a) and the
deviations in table \ref{ce_vib}(b).
Although only few clusters were used in this expansion, the convergence is
really good.

\begin{table}[!hbtp]

\caption{Cluster expansion of the function $f^s=3\ln\theta_D-1$ for the
vibrational free energy.}
\label{ce_vib}

\begin{minipage}{0.45\linewidth}
\subfigure[Coefficients of the expansion.]{
\begin{ruledtabular}
\begin{tabular}{c.{0}.{3}}
Cluster & \multicolumn{1}{c}{$D_{\alpha}$}      &
\multicolumn{1}{c}{$J_{\alpha}$}\\
\hline
\{0\}   & 1     & 17.385        \\
\{1\}   & 1     & 0.874         \\
\{2,1\} & 6     & -0.197        \\
\{3,1\} & 8     & -0.027        \\
\end{tabular}
\end{ruledtabular}
}
\end{minipage}
\hfill
\begin{minipage}{0.45\linewidth}
\subfigure[Deviation $\delta f^s$ of the expansion.]{
\begin{ruledtabular}
\begin{tabular}{l.{2}.{2}}
& \multicolumn{1}{c}{$f^s$}& \multicolumn{1}{c}{$\delta f^s$}\\
\hline
Al (fcc)                & 16.86 & 0.    \\
Al$_7$Zr (D1)           & 17.32 & -0.08 \\
Al$_3$Zr (L1$_2$)       & 17.97 & 0.04  \\
Zr$_3$Al (L1$_2$)       & 16.88 & 0.04  \\
Zr$_7$Al (D1)           & 16.11 & -0.08 \\
Zr (fcc)                & 15.55 & 0.    \\
\end{tabular}
\end{ruledtabular}
}
\end{minipage}

\end{table}

%%%%%%%%%%%%%%%%%%%%%%%%%%%%%%%%%%%%%%%%%%%%%%%%%%

\subsection{Bragg-Williams approximation}

We thus obtained an expression for the different parts of the free energy
functional $F[\rho]$ of expression (\ref{functionalF}): the cohesive part is given by the cluster expansion
of the FP-LMTO calculations (coefficients in table \ref{coef_ce_total}),
the vibrational energy by the expression (\ref{Fvib_ce}) with the coefficients
of table \ref{ce_vib}(a), and the electronic contribution can be neglected.
The functional $F[\rho]$ is minimized in the Bragg-Williams approximation.
This assumes that there is no short range order and that the correlation
functions can be factorized over the mean values of the pseudo spin variable
$\langle\sigma_n\rangle$ for the lattice sites contained in the cluster,
\begin{equation}
\zeta_{\alpha} = \langle \prod_{i\in\alpha} \sigma_i \rangle
= \prod_{i\in\alpha} \langle \sigma_i \rangle .
\end{equation}
The Bragg-Williams approximation thus assumes that the lattice sites
interact only through their mean occupancy and neglects all correlations
between different sites.
This can be improved by using the Cluster Variation Method (CVM)\cite{KIK51},
but in the case of a low solubility, no really important improvement is
expected when going from the Bragg-Williams approximation to the CVM.
Moreover, the computational time necessary to obtain the free energy by means of the
CVM increases a lot with the size of the maximal cluster.
As Zr has a really low solubility in Al~(fcc) and as the long range
interactions of the cluster expansion of the formation energy
requires a too large cluster, we chose to work with the Bragg-Williams
approximation.

Within the Bragg-Williams approximation, the configurational entropy has the
following expression for a binary compound
\begin{eqnarray}
S[\rho] &=& - k_B \sum_n ( 1+\langle\sigma_n\rangle)
\ln(1+\langle\sigma_n\rangle) \nonumber\\
&&+ (1-\langle\sigma_n\rangle) \ln(1-\langle\sigma_n\rangle)
\label{BW_entropy} .
\end{eqnarray}

%%%%%%%%%%%%%%%%%%%%%%%%%%%%%%%%%
\subsubsection{Disordered phase}

For a disordered state, all lattice sites are equivalent by symmetry.
They have thus the same point correlation $\zeta_1=2x-1$, where $x$ is the
Zr atomic concentration. Consequently any correlation function can be written
in terms of the point correlation:
\begin{equation}
\zeta_{\alpha} = \zeta_1^{|\alpha|} .
\end{equation}

The cluster expansion of the function $f^s$, using the expression
(\ref{ce_excess}) of the excess function $\Delta f^s$, can then be
expressed as a function of the point correlation, or equally as a function
of the concentration.
This leads to an expression similar to the way
the internal energy of a solid solution is written in a Redlich-Kister
model which is of common use in the Calphad method\cite{CALPHAD}
\begin{equation}
f^s = x f^A + (1-x) f^B + x(1-x) \sum_{n\ge0} L_n (2x-1)^n
\label{RK} ,
\end{equation}
where the coefficients $L_n$ are obtained from the coefficients $J_{\alpha}$
by the relations
\begin{equation}
L_n=-4 \sum_{i\ge1}  \sum_{\substack{\alpha \\ |\alpha|=n+2i}}
D_{\alpha}f_{\alpha} .
\end{equation}
Using the expression (\ref{BW_entropy}) for the entropy,
we obtain for the free energy of the disordered fcc solid solution
Al$_{(1-x)}$Zr$_x$
\begin{eqnarray}
F(x) &=& (1-x) U^{Al,fcc}  + x U^{Zr,fcc} \nonumber\\
&&+ k_B T \left[ x \ln x + (1-x)\ln(1-x) \right] \nonumber\\
&&+ x(1-x) \sum_{n\ge0} L_n (2x-1)^n
\label{RK_free_energy} .
\end{eqnarray}

The Redlich-Kister coefficients are obtained from the cluster expansion
of the formation energy  and the
cluster expansion of the vibrational free energy calculated in the Debye approximation.
\begin{eqnarray}
L_0 &=& -89.09 + 29.9\times10^{-3} T \textrm{ mRy/atom}
\nonumber\\
L_1 &=& -14.30 + 5.47\times10^{-3} T \textrm{ mRy/atom}
\label{RK_relaxed}\\
L_2 &=& -7.03 \textrm{ mRy/atom}
\nonumber
\end{eqnarray}

For a dilute solution ($x\ll1$), the expression (\ref{RK_free_energy}) is equivalent to the free energy
of a regular solution, the excess free energy being then $x(1-x)\Omega=x(1-x)(L_0+L_1+L_2)$.
We compare in table \ref{excess_table} the value of $\Omega$ obtained from our calculations
to the ones obtained by a fit of the phase diagram through a Calphad approach \cite{SAU89,MUR92}.

\begin{table}[!hbt]

\caption{Parameter $\Omega$ (in mRy/atom) of the excess free energy for an fcc regular solid solution Al-Zr
deduced from \abinitio calculations
and compared to values obtained by a fit of the experimental phase diagram.}
\label{excess_table}

\begin{ruledtabular}
\begin{tabular}{lccl}
Present work    & $\Omega$&=& $-110.42+35.37\times10^{-3}T$\\
Saunders \cite{SAU89}      &&=& $-87.60+22.85\times10^{-3}T$\\
Murray \etal \cite{MUR92} &&=& $-85.08+31.01\times10^{-3}T$ \\
\end{tabular}
\end{ruledtabular}

\end{table}

%%%%%%%%%%%%%%%%%%%%%%%%%%%%%%%%%%%
\subsubsection{Line compounds}

Al$_3$Zr in the DO$_{23}$ or L1$_2$ structures can be considered as a line
compound, \ie perfectly ordered:
both structures are composed of interpenetrating sublattices of
pure Al and pure Zr.
The configurational entropy of such line compounds can be neglected and
these structures only exist for a concentration $x=1/4$.
We checked with a calculation using the previous cluster expansions of the formation
and vibrational energies that this assumption was correct in the range of temperature of interest
when looking at equilibrium with the solid solution.
The free energy of these compounds is then simply given by
\begin{equation}
F^{Al_3Zr} = \frac{3}{4} U^{Al,fcc} + \frac{1}{4} U^{Zr,fcc}
+ \Delta U^{Al_3Zr} ,
\end{equation}
where $\Delta U^{Al_3Zr}$ (in mRy/atom) is obtained from our previous calculations of the
formation energy (Table \ref{AlZr_relaxed_lda}) and of the excess vibrational free energy
calculated from the phonon DOS (Table \ref{highT_table}),
\begin{subequations}\begin{eqnarray}
\Delta U^{Al_3Zr,L1_2} &=& -39.00 + 5.38\times10^{-3} T ,\\
\Delta U^{Al_3Zr,DO_{23}} &=& -40.72 + 5.38\times10^{-3} T .
\end{eqnarray}\end{subequations}

%%%%%%%%%%%%%%%%%%%%%%%%%%%%%%%%%%%%%%%%%%%%%%%%%%%%%%%%%%%%%%%%%%%%%%%%%%%%%%%
\subsection{Solubility limit of Zr in Al~(fcc)}

Using the previous expressions for the free energies of the disordered phase
and the line compounds Al$_3$Zr, we obtained the solubility limit of Zr in
Al~(fcc), both in the stable phase diagram when considering the structure
DO$_{23}$ for Al$_3$Zr and in the metastable one when considering
the structure L1$_2$.
As we are in the case of a dilute solid solution, the solubility limit of Zr in Al~(fcc) is
an analytic function of the temperature\cite{SIG98},
\begin{equation}
x=\exp\left( \frac{4\Delta U^{Al_3Zr}-\Omega}{k_BT} \right) .
\end{equation}

\begin{figure}[!hbtp]
\includegraphics[width=0.45\textwidth]{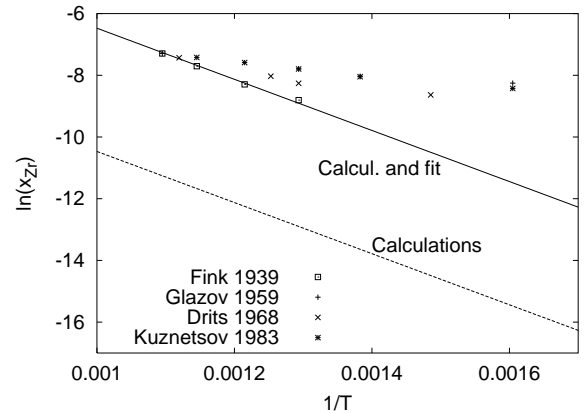}
\caption{Fit of the entropy of the solid solution so as to reproduce Fink experimental data.}
\label{fit}
\end{figure}

The solubility we obtained is too low: at the melting temperature of the perictectic it is equal to
0.0016~at.\%~Zr, whereas the one deduced from experimental data is
0.08~at.\%~Zr \cite{MUR92}.
When comparing the variation with respect to $1/T$ of $\ln x$ with experimental measurements,
we obtain a straight line having the same slope as Fink data \cite{FIN39}  (\cf Fig. \ref{fit}).
This shows that our calculations provide an approximation of the enthalpy difference between
the solid solution and the DO$_{23}$ ordered compound which is consistent with Finks data,
and that the discrepancy on the solubility limit only arises from an error on the estimation of
the entropy difference.
Computing the solubility limit of Sc in Al, Asta \etal\cite{AST98} reached the same conclusion
that \abinitio calculations correctly predicted the enthalpy difference  between the ordered compound
and the solid solution when compared to experimental data.
In our case, the error on the entropic terms may come from an overestimation of the vibrational free energy of the disordered phase
due to the use of the Debye model for this phase.
As for the structures DO$_{23}$ and L1$_2$ of Al$_3$Zr, the Debye model overestimates
the excess vibrational free energy by a factor $\sim2$ (Table \ref{highT_table}),
we can think that we get an error of the same range for the solid solution.

We correct the entropic part, leaving unchanged the enthalpic part,
of the parameter $\Omega$ defining the excess free energy of the solid solution
so as to obtain a perfect agreement with Fink data (\cf Fig. \ref{fit}), and we obtain
\begin{equation}
\Omega = -110.42+10.07\times10^{-3}T \textrm{~mRy/atom.}
\label{omega_corrected}
\end{equation}
We thus get a stable solubility limit that is consistent with Fink measurements, and we are now able
to predict the metastable limit using the expression (\ref{omega_corrected}) to evaluate
the excess free energy of the solid solution.
As the structures DO$_{23}$ and L1$_2$ of Al$_3$Zr have
the same vibrational free energy, the difference of solubility limit is only due
to the difference of ground state energies of these two phases.
At the melting temperature of the peritectic, we obtain a maximal metastable
solubility limit equal to 0.275~at.\%~Zr.

% But this approximation was necessary to make the problem tractable computationally.
% Another solution to obtain the vibrational free energy of the disordered solid solution
% would have been to calculate the phonon spectrum of a supercell with a low Zr concentration,
% like Al$_{31}$Zr, and to equal the vibrational free energy of this structure with the one of the
% solid solution as Ozoli\c{n}\v{s} and Asta\cite{OZO01} did for the Al-Sc system.
% Nevertheless, when comparing our results with experimental data,
% one should remember that no fitting procedure was used and that all the
% parameters of the free energies used in the thermodynamic model were
% obtained from first principles calculations.
% Moreover, the solubility limit of Zr in Al~(fcc) is really low and really
% sensitive to a little change in the value of the free energy.

\begin{figure}[!hbtp]
\includegraphics[width=0.45\textwidth]{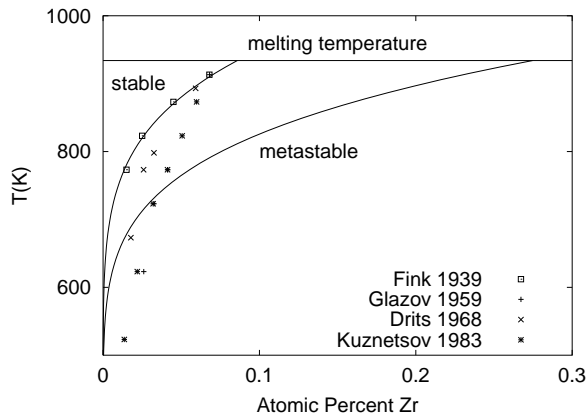}
\caption[stable and metastable solubility limits of Zr in Al]
{Calculated stable and metastable solubility limits of Zr in Al
compared to experimental data\cite{FIN39, GLA59, DRI68, KUZ83} }
\label{solubility_relaxed}
\end{figure}

This is to compare to the results obtained by a Calphad method. Murray \etal
\cite{MUR92} modelled the metastable phase of Al$_3$Zr as a line compound.
They assumed that only an enthalpy term, and no entropy term, contributes
to the free energy difference between the stable and metastable phases.
This was done to guarantee that L1$_2$ does not become stable at high
temperature.
Moreover there is no experimental data that allows to estimate the entropy
of the L1$_2$ phase.
Our calculation of the vibrational free energy shows that such an
approximation was correct.
The enthalpy difference between the two compounds was assumed to arise from
the coherency of the phase L1$_2$ with the matrix. They calculated from the
elastic properties of Al and an estimate of the composition dependence of
the lattice parameter an elastic energy of 1.52~mRy/atom. This estimation
is quite close to our calculation ($\Delta H=1.72$~mRy/atom)
as well as to the experimental measurement of Desch \etal \cite{DES91}
($\Delta H=1.69$~mRy/atom).
They thus obtained a solubility limit that is higher in the metastable phase
diagram than in the stable one and their prevision is really close to our result:
they predicted a maximal metastable solubility limit equal to 0.21~at.\%.

In another Calphad study, Saunders \cite{SAU89} used the Gibbs energy for
the disordered Al~(fcc) solution, as derived from the stable equilibrium
diagram, to construct the Gibbs energy of the ordered L1$_2$ phase in the
Bragg-Williams approximation. He found a higher solubility limit for
Zr in the metastable phase diagram than Murray \etal as he predicted a
metastable solvus composition of 0.3~at.\%~Zr at the melting temperature of the peritectic.

Our study thus allows one to estimate the free energy difference
between the stable and metastable phases of Al$_3$Zr,
quantity which is not available experimentally and has to be guessed in these
Calphad studies.
One thus sees how it is possible to improve the thermodynamic database
available to Calphad methods.

%%%%%%%%%%%%%%%%%%%%%%%%%%%%%%%%%%%%%%%%%%%

\section{Conclusion}

The equation of state for several compounds in the Al-Zr system
has been computed using the full potential linear muffin tin orbital method
(FP-LMTO).
These \abinitio calculations correctly predict the stability of the phase
DO$_{23}$ for Al$_3$Zr if we consider the cell internal relaxations.

We made a cluster expansion of the results of \abinitio calculations
to predict the formation energy of any compound in the Al-Zr system
based on an fcc underlying lattice.
We showed that despite the size difference between Al and Zr
a totally or globally relaxed expansion for the volume leads to
the same result: there is no difference if we use the cluster expansion to
predict the formation energy at the equilibrium volume of each structure or
at a fixed volume, the energy being then minimized according to the volume.

For finite temperature calculations, we showed that the electronic
excitations can be neglected.
The vibrational energy was studied in the harmonic model, using different
levels of approximation: the Debye model was compared to results obtained
from a calculation of the phonon spectrum for Al$_3$Zr in the structures
DO$_{23}$  and L1$_2$, and
it was found that the use of the Debye model leads to an overestimation of
the vibrational free energy. So we preferred to use the results from the
phonon spectrum to calculate the vibrational free energy of ordered compounds.
For the disordered phase, we chose to make a cluster expansion of the vibrational free energy.
It was only possible with the Debye model as this requires less computational time.

We were able to calculate the solubility limit of Zr in Al~(fcc) in the the Bragg-Williams approximation.
The solubility limit obtained is too low compared to experimental data.
We showed that this discrepancy is due to an error in the estimation of the entropy
in our thermodynamic model.
This may be due to an overestimation of the vibrational free energy of the disordered phase
because of the use of the Debye model for this phase.
Correcting the vibrational entropy of the solid solution so as to fit the experimental measurements
of Fink, we were able to predict the metastable solubility limit which lies between the
estimation of Murray and the one of Saunders, both obtained by a Calphad method.
We thus showed how first principles calculations can lead to an estimation
of the phase diagram.
This approach has the advantage of not requiring any experimental input, and consequently
this is not a problem to predict stability of metastable phases.

%%%%%%%%%%%%%%%%%%%%%%%%%%%%%%%%%%%%%%%%%%%%%%%%%%%%

\begin{acknowledgments}
The authors would like to thank M. Nastar for valuable discussions.
Financial support from Pechiney~CRV (France) is acknowledged.
\end{acknowledgments}

%%%%%%%%%%%%%%%%%%%%%%%%%%%%%%%%%%%%%%%%%%%%%%%%%

\bibliographystyle{apsrev}
%\bibliography{biblio}

\begin{thebibliography}{54}
\expandafter\ifx\csname natexlab\endcsname\relax\def\natexlab#1{#1}\fi
\expandafter\ifx\csname bibnamefont\endcsname\relax
  \def\bibnamefont#1{#1}\fi
\expandafter\ifx\csname bibfnamefont\endcsname\relax
  \def\bibfnamefont#1{#1}\fi
\expandafter\ifx\csname citenamefont\endcsname\relax
  \def\citenamefont#1{#1}\fi
\expandafter\ifx\csname url\endcsname\relax
  \def\url#1{\texttt{#1}}\fi
\expandafter\ifx\csname urlprefix\endcsname\relax\def\urlprefix{URL }\fi
\providecommand{\bibinfo}[2]{#2}
\providecommand{\eprint}[2][]{\url{#2}}

\bibitem[{\citenamefont{Hohenberg and Kohn}(1964)}]{HOH64}
\bibinfo{author}{\bibfnamefont{P.}~\bibnamefont{Hohenberg}} \bibnamefont{and}
  \bibinfo{author}{\bibfnamefont{W.}~\bibnamefont{Kohn}},
  \bibinfo{journal}{Phys. Rev.} \textbf{\bibinfo{volume}{136}},
  \bibinfo{pages}{B864} (\bibinfo{year}{1964}).

\bibitem[{\citenamefont{Kohn and Sham}(1965)}]{KOH65}
\bibinfo{author}{\bibfnamefont{W.}~\bibnamefont{Kohn}} \bibnamefont{and}
  \bibinfo{author}{\bibfnamefont{L.~J.} \bibnamefont{Sham}},
  \bibinfo{journal}{Phys. Rev.} \textbf{\bibinfo{volume}{140}},
  \bibinfo{pages}{A1133} (\bibinfo{year}{1965}).

\bibitem[{\citenamefont{Ducastelle}(1991)}]{DUCASTELLE}
\bibinfo{author}{\bibfnamefont{F.}~\bibnamefont{Ducastelle}},
  \emph{\bibinfo{title}{Order and Phase Stability in Alloys}}
  (\bibinfo{publisher}{North-Holland Amsterdam}, \bibinfo{year}{1991}).

\bibitem[{\citenamefont{de~Fontaine}(1994)}]{DEF94}
\bibinfo{author}{\bibfnamefont{D.}~\bibnamefont{de~Fontaine}},
  \bibinfo{journal}{Solid State Physics} \textbf{\bibinfo{volume}{47}},
  \bibinfo{pages}{33} (\bibinfo{year}{1994}).

\bibitem[{\citenamefont{Wolverton and Zunger}(1995)}]{WOL95}
\bibinfo{author}{\bibfnamefont{C.}~\bibnamefont{Wolverton}} \bibnamefont{and}
  \bibinfo{author}{\bibfnamefont{A.}~\bibnamefont{Zunger}},
  \bibinfo{journal}{Phys. Rev. B} \textbf{\bibinfo{volume}{52}},
  \bibinfo{pages}{8813} (\bibinfo{year}{1995}).

\bibitem[{\citenamefont{Garbulsky and Ceder}(1994)}]{GAR94}
\bibinfo{author}{\bibfnamefont{G.~D.} \bibnamefont{Garbulsky}}
  \bibnamefont{and} \bibinfo{author}{\bibfnamefont{G.}~\bibnamefont{Ceder}},
  \bibinfo{journal}{Phys. Rev. B} \textbf{\bibinfo{volume}{49}},
  \bibinfo{pages}{6327} (\bibinfo{year}{1994}).

\bibitem[{\citenamefont{Ozoli\c{n}\v{s} and Asta}(2001)}]{OZO01}
\bibinfo{author}{\bibfnamefont{V.}~\bibnamefont{Ozoli\c{n}\v{s}}}
  \bibnamefont{and} \bibinfo{author}{\bibfnamefont{M.}~\bibnamefont{Asta}},
  \bibinfo{journal}{Phys. Rev. Lett.} \textbf{\bibinfo{volume}{86}},
  \bibinfo{pages}{448} (\bibinfo{year}{2001}).

\bibitem[{\citenamefont{Fink and Willey}(1939)}]{FIN39}
\bibinfo{author}{\bibfnamefont{W.~L.} \bibnamefont{Fink}} \bibnamefont{and}
  \bibinfo{author}{\bibfnamefont{L.~A.} \bibnamefont{Willey}},
  \bibinfo{journal}{Trans. AIME} \textbf{\bibinfo{volume}{133}},
  \bibinfo{pages}{69} (\bibinfo{year}{1939}).

\bibitem[{\citenamefont{Glazov et~al.}(1959)\citenamefont{Glazov, Lazarev, and
  Korolkov}}]{GLA59}
\bibinfo{author}{\bibfnamefont{V.~M.} \bibnamefont{Glazov}},
  \bibinfo{author}{\bibfnamefont{G.}~\bibnamefont{Lazarev}}, \bibnamefont{and}
  \bibinfo{author}{\bibfnamefont{G.}~\bibnamefont{Korolkov}},
  \bibinfo{journal}{Metallovedenie i Termicheskaia Obrabotka Metallov.}
  \textbf{\bibinfo{volume}{10}}, \bibinfo{pages}{48} (\bibinfo{year}{1959}).

\bibitem[{\citenamefont{Drits et~al.}(1968)\citenamefont{Drits, Kadaner, and
  Kuz'mina}}]{DRI68}
\bibinfo{author}{\bibfnamefont{M.~E.} \bibnamefont{Drits}},
  \bibinfo{author}{\bibfnamefont{Y.~S.} \bibnamefont{Kadaner}},
  \bibnamefont{and} \bibinfo{author}{\bibfnamefont{V.~I.}
  \bibnamefont{Kuz'mina}}, \bibinfo{journal}{Izv. Akad. Nauk {SSSR} Met.}
  \textbf{\bibinfo{volume}{1}}, \bibinfo{pages}{102} (\bibinfo{year}{1968}).

\bibitem[{\citenamefont{Kuznetsov et~al.}(1983)\citenamefont{Kuznetsov,
  Barsukov, and Abas}}]{KUZ83}
\bibinfo{author}{\bibfnamefont{G.~M.} \bibnamefont{Kuznetsov}},
  \bibinfo{author}{\bibfnamefont{A.~V.} \bibnamefont{Barsukov}},
  \bibnamefont{and} \bibinfo{author}{\bibfnamefont{M.~I.} \bibnamefont{Abas}},
  \bibinfo{journal}{Sov. Non-Ferrous Met. Res.} \textbf{\bibinfo{volume}{11}},
  \bibinfo{pages}{47} (\bibinfo{year}{1983}).

\bibitem[{\citenamefont{Ryum}(1969)}]{RYU69}
\bibinfo{author}{\bibfnamefont{N.}~\bibnamefont{Ryum}}, \bibinfo{journal}{Acta
  Met.} \textbf{\bibinfo{volume}{17}}, \bibinfo{pages}{269}
  (\bibinfo{year}{1969}).

\bibitem[{\citenamefont{Nes}(1972)}]{NES72}
\bibinfo{author}{\bibfnamefont{E.}~\bibnamefont{Nes}}, \bibinfo{journal}{Acta
  Met.} \textbf{\bibinfo{volume}{20}}, \bibinfo{pages}{499}
  (\bibinfo{year}{1972}).

\bibitem[{\citenamefont{Nes and Billdal}(1977)}]{NES77}
\bibinfo{author}{\bibfnamefont{E.}~\bibnamefont{Nes}} \bibnamefont{and}
  \bibinfo{author}{\bibfnamefont{H.}~\bibnamefont{Billdal}},
  \bibinfo{journal}{Acta Met.} \textbf{\bibinfo{volume}{25}},
  \bibinfo{pages}{1031} (\bibinfo{year}{1977}).

\bibitem[{\citenamefont{Andersen}(1975)}]{AND75}
\bibinfo{author}{\bibfnamefont{O.~K.} \bibnamefont{Andersen}},
  \bibinfo{journal}{Phys. Rev. B} \textbf{\bibinfo{volume}{12}},
  \bibinfo{pages}{3060} (\bibinfo{year}{1975}).

\bibitem[{\citenamefont{Methfessel}(1988)}]{MET88}
\bibinfo{author}{\bibfnamefont{M.}~\bibnamefont{Methfessel}},
  \bibinfo{journal}{Phys. Rev. B} \textbf{\bibinfo{volume}{38}},
  \bibinfo{pages}{1537} (\bibinfo{year}{1988}).

\bibitem[{\citenamefont{Methfessel et~al.}(1989)\citenamefont{Methfessel,
  Rodriguez, and Andersen}}]{MET89}
\bibinfo{author}{\bibfnamefont{M.}~\bibnamefont{Methfessel}},
  \bibinfo{author}{\bibfnamefont{C.~O.} \bibnamefont{Rodriguez}},
  \bibnamefont{and} \bibinfo{author}{\bibfnamefont{O.~K.}
  \bibnamefont{Andersen}}, \bibinfo{journal}{Phys. Rev. B}
  \textbf{\bibinfo{volume}{40}}, \bibinfo{pages}{2009} (\bibinfo{year}{1989}).

\bibitem[{\citenamefont{Methfessel and van Schilfgaarde}(1993)}]{MET93}
\bibinfo{author}{\bibfnamefont{M.}~\bibnamefont{Methfessel}} \bibnamefont{and}
  \bibinfo{author}{\bibfnamefont{M.}~\bibnamefont{van Schilfgaarde}},
  \bibinfo{journal}{Phys. Rev. B} \textbf{\bibinfo{volume}{48}},
  \bibinfo{pages}{4937} (\bibinfo{year}{1993}).

\bibitem[{\citenamefont{von Barth and Hedin}(1972)}]{VON72}
\bibinfo{author}{\bibfnamefont{U.}~\bibnamefont{von Barth}} \bibnamefont{and}
  \bibinfo{author}{\bibfnamefont{L.}~\bibnamefont{Hedin}}, \bibinfo{journal}{J.
  Phys. C} \textbf{\bibinfo{volume}{5}}, \bibinfo{pages}{1629}
  (\bibinfo{year}{1972}).

\bibitem[{\citenamefont{Jomard et~al.}(1998)\citenamefont{Jomard, Magaud, and
  Pasturel}}]{JOM98}
\bibinfo{author}{\bibfnamefont{G.}~\bibnamefont{Jomard}},
  \bibinfo{author}{\bibfnamefont{L.}~\bibnamefont{Magaud}}, \bibnamefont{and}
  \bibinfo{author}{\bibfnamefont{A.}~\bibnamefont{Pasturel}},
  \bibinfo{journal}{Philos. Mag. B} \textbf{\bibinfo{volume}{77}},
  \bibinfo{pages}{67} (\bibinfo{year}{1998}).

\bibitem[{\citenamefont{Rose et~al.}(1984)\citenamefont{Rose, Smith, Guinea,
  and Ferrante}}]{ROS84}
\bibinfo{author}{\bibfnamefont{J.~H.} \bibnamefont{Rose}},
  \bibinfo{author}{\bibfnamefont{J.~R.} \bibnamefont{Smith}},
  \bibinfo{author}{\bibfnamefont{F.}~\bibnamefont{Guinea}}, \bibnamefont{and}
  \bibinfo{author}{\bibfnamefont{J.}~\bibnamefont{Ferrante}},
  \bibinfo{journal}{Phys. Rev. B} \textbf{\bibinfo{volume}{29}},
  \bibinfo{pages}{2963} (\bibinfo{year}{1984}).

\bibitem[{\citenamefont{Amador et~al.}(1995)\citenamefont{Amador, Hoyt,
  Chakoumakos, and de~Fontaine}}]{AMA95}
\bibinfo{author}{\bibfnamefont{C.}~\bibnamefont{Amador}},
  \bibinfo{author}{\bibfnamefont{J.~J.} \bibnamefont{Hoyt}},
  \bibinfo{author}{\bibfnamefont{B.~C.} \bibnamefont{Chakoumakos}},
  \bibnamefont{and}
  \bibinfo{author}{\bibfnamefont{D.}~\bibnamefont{de~Fontaine}},
  \bibinfo{journal}{Phys. Rev. Lett.} \textbf{\bibinfo{volume}{74}},
  \bibinfo{pages}{4955} (\bibinfo{year}{1995}).

\bibitem[{\citenamefont{Colinet and Pasturel}(2001)}]{COL01}
\bibinfo{author}{\bibfnamefont{C.}~\bibnamefont{Colinet}} \bibnamefont{and}
  \bibinfo{author}{\bibfnamefont{A.}~\bibnamefont{Pasturel}},
  \bibinfo{journal}{J. Alloys Comp.} \textbf{\bibinfo{volume}{319}},
  \bibinfo{pages}{154} (\bibinfo{year}{2001}).

\bibitem[{\citenamefont{Desch et~al.}(1991)\citenamefont{Desch, Schwarz, and
  Nash}}]{DES91}
\bibinfo{author}{\bibfnamefont{P.~B.} \bibnamefont{Desch}},
  \bibinfo{author}{\bibfnamefont{R.~B.} \bibnamefont{Schwarz}},
  \bibnamefont{and} \bibinfo{author}{\bibfnamefont{P.}~\bibnamefont{Nash}},
  \bibinfo{journal}{J. Less-Common Metals} \textbf{\bibinfo{volume}{168}},
  \bibinfo{pages}{69} (\bibinfo{year}{1991}).

\bibitem[{\citenamefont{Sanchez et~al.}(1984)\citenamefont{Sanchez, Ducastelle,
  and Gratias}}]{SAN84}
\bibinfo{author}{\bibfnamefont{J.~M.} \bibnamefont{Sanchez}},
  \bibinfo{author}{\bibfnamefont{F.}~\bibnamefont{Ducastelle}},
  \bibnamefont{and} \bibinfo{author}{\bibfnamefont{D.}~\bibnamefont{Gratias}},
  \bibinfo{journal}{Physica} \textbf{\bibinfo{volume}{128A}},
  \bibinfo{pages}{334} (\bibinfo{year}{1984}).

\bibitem[{\citenamefont{Connolly and Williams}(1983)}]{CON83}
\bibinfo{author}{\bibfnamefont{J.~W.} \bibnamefont{Connolly}} \bibnamefont{and}
  \bibinfo{author}{\bibfnamefont{A.~R.} \bibnamefont{Williams}},
  \bibinfo{journal}{Phys. Rev. B} \textbf{\bibinfo{volume}{27}},
  \bibinfo{pages}{5169} (\bibinfo{year}{1983}).

\bibitem[{\citenamefont{Carlsson}(1987)}]{CAR87}
\bibinfo{author}{\bibfnamefont{A.~E.} \bibnamefont{Carlsson}},
  \bibinfo{journal}{Phys. Rev. B} \textbf{\bibinfo{volume}{35}},
  \bibinfo{pages}{4858} (\bibinfo{year}{1987}).

\bibitem[{\citenamefont{Sanchez}(1992)}]{SAN92}
\bibinfo{author}{\bibfnamefont{J.~M.} \bibnamefont{Sanchez}}, in
  \emph{\bibinfo{booktitle}{Structural and Phase Stability of Alloys}}, edited
  by \bibinfo{editor}{\bibfnamefont{J.~L.} \bibnamefont{Mor\'an-L\'opez}},
  \bibinfo{editor}{\bibfnamefont{F.}~\bibnamefont{Mej\'ia-Lira}},
  \bibnamefont{and} \bibinfo{editor}{\bibfnamefont{J.~M.}
  \bibnamefont{Sanchez}} (\bibinfo{year}{1992}), pp. \bibinfo{pages}{151--165}.

\bibitem[{\citenamefont{Baroni et~al.}(1987)\citenamefont{Baroni, Giannozzi,
  and Testa}}]{BAR87}
\bibinfo{author}{\bibfnamefont{S.}~\bibnamefont{Baroni}},
  \bibinfo{author}{\bibfnamefont{P.}~\bibnamefont{Giannozzi}},
  \bibnamefont{and} \bibinfo{author}{\bibfnamefont{A.}~\bibnamefont{Testa}},
  \bibinfo{journal}{Phys. Rev. Lett.} \textbf{\bibinfo{volume}{58}},
  \bibinfo{pages}{1861} (\bibinfo{year}{1987}).

\bibitem[{\citenamefont{Savrasov}(1992)}]{SAV92b}
\bibinfo{author}{\bibfnamefont{S.~Y.} \bibnamefont{Savrasov}},
  \bibinfo{journal}{Phys. Rev. Lett.} \textbf{\bibinfo{volume}{69}},
  \bibinfo{pages}{2819} (\bibinfo{year}{1992}).

\bibitem[{\citenamefont{Savrasov}(1996)}]{SAV96}
\bibinfo{author}{\bibfnamefont{S.~Y.} \bibnamefont{Savrasov}},
  \bibinfo{journal}{Phys. Rev. B} \textbf{\bibinfo{volume}{54}},
  \bibinfo{pages}{16470} (\bibinfo{year}{1996}).

\bibitem[{\citenamefont{Moruzzi et~al.}(1978)\citenamefont{Moruzzi, Janak, and
  Williams}}]{MOR78}
\bibinfo{author}{\bibfnamefont{V.~L.} \bibnamefont{Moruzzi}},
  \bibinfo{author}{\bibfnamefont{J.~F.} \bibnamefont{Janak}}, \bibnamefont{and}
  \bibinfo{author}{\bibfnamefont{A.~R.} \bibnamefont{Williams}},
  \emph{\bibinfo{title}{Calculated Electronic Properties of Metals}}
  (\bibinfo{publisher}{Pergamon}, \bibinfo{address}{New York},
  \bibinfo{year}{1978}).

\bibitem[{\citenamefont{Stedman and Nilsson}(1966)}]{STE66}
\bibinfo{author}{\bibfnamefont{R.}~\bibnamefont{Stedman}} \bibnamefont{and}
  \bibinfo{author}{\bibfnamefont{G.}~\bibnamefont{Nilsson}},
  \bibinfo{journal}{Phys. Rev.} \textbf{\bibinfo{volume}{145}},
  \bibinfo{pages}{492} (\bibinfo{year}{1966}).

\bibitem[{\citenamefont{Dederichs et~al.}(1981)\citenamefont{Dederichs,
  Schober, and Sellmyer}}]{LANDOLT2}
\bibinfo{author}{\bibfnamefont{P.}~\bibnamefont{Dederichs}},
  \bibinfo{author}{\bibfnamefont{H.}~\bibnamefont{Schober}}, \bibnamefont{and}
  \bibinfo{author}{\bibfnamefont{D.}~\bibnamefont{Sellmyer}}, in
  \emph{\bibinfo{booktitle}{{L}andolt-{B}\"ornstein}}, edited by
  \bibinfo{editor}{\bibfnamefont{K.-H.} \bibnamefont{Hellwege}}
  \bibnamefont{and} \bibinfo{editor}{\bibfnamefont{O.}~\bibnamefont{Madelung}}
  (\bibinfo{publisher}{Springer}, \bibinfo{address}{Berlin},
  \bibinfo{year}{1981}), vol. \bibinfo{volume}{{III}/13(a)}.

\bibitem[{\citenamefont{Quong and Klein}(1992)}]{QUO92}
\bibinfo{author}{\bibfnamefont{A.~A.} \bibnamefont{Quong}} \bibnamefont{and}
  \bibinfo{author}{\bibfnamefont{B.~M.} \bibnamefont{Klein}},
  \bibinfo{journal}{Phys. Rev. B} \textbf{\bibinfo{volume}{46}},
  \bibinfo{pages}{10734} (\bibinfo{year}{1992}).

\bibitem[{\citenamefont{de~Gironcoli}(1995)}]{GIR95}
\bibinfo{author}{\bibfnamefont{S.}~\bibnamefont{de~Gironcoli}},
  \bibinfo{journal}{Phys. Rev. B} \textbf{\bibinfo{volume}{51}},
  \bibinfo{pages}{6773} (\bibinfo{year}{1995}).

\bibitem[{\citenamefont{Bauer et~al.}(1998)\citenamefont{Bauer, Schmid, Pavone,
  and Strauch}}]{BAU98}
\bibinfo{author}{\bibfnamefont{R.}~\bibnamefont{Bauer}},
  \bibinfo{author}{\bibfnamefont{A.}~\bibnamefont{Schmid}},
  \bibinfo{author}{\bibfnamefont{P.}~\bibnamefont{Pavone}}, \bibnamefont{and}
  \bibinfo{author}{\bibfnamefont{D.}~\bibnamefont{Strauch}},
  \bibinfo{journal}{Phys. Rev. B} \textbf{\bibinfo{volume}{57}},
  \bibinfo{pages}{11276} (\bibinfo{year}{1998}).

\bibitem[{\citenamefont{Cowley}(1974)}]{COW74}
\bibinfo{author}{\bibfnamefont{E.~R.} \bibnamefont{Cowley}},
  \bibinfo{journal}{Can. J. Phys.} \textbf{\bibinfo{volume}{52}},
  \bibinfo{pages}{1714} (\bibinfo{year}{1974}).

\bibitem[{\citenamefont{Gilat and Nicklow}(1966)}]{GIL66}
\bibinfo{author}{\bibfnamefont{G.}~\bibnamefont{Gilat}} \bibnamefont{and}
  \bibinfo{author}{\bibfnamefont{R.~M.} \bibnamefont{Nicklow}},
  \bibinfo{journal}{Phys. Rev.} \textbf{\bibinfo{volume}{143}},
  \bibinfo{pages}{487} (\bibinfo{year}{1966}).

\bibitem[{\citenamefont{Alers}(1965)}]{ALE65}
\bibinfo{author}{\bibfnamefont{G.~A.} \bibnamefont{Alers}}, in
  \emph{\bibinfo{booktitle}{Physical Acoustics}}, edited by
  \bibinfo{editor}{\bibfnamefont{W.~P.} \bibnamefont{Mason}}
  (\bibinfo{publisher}{Academic}, \bibinfo{address}{New York},
  \bibinfo{year}{1965}), vol. \bibinfo{volume}{III-B}, pp.
  \bibinfo{pages}{1--42}.

\bibitem[{\citenamefont{S\"oderlind et~al.}(1993)\citenamefont{S\"oderlind,
  Eriksson, Wills, and Boring}}]{SOD93}
\bibinfo{author}{\bibfnamefont{P.}~\bibnamefont{S\"oderlind}},
  \bibinfo{author}{\bibfnamefont{O.}~\bibnamefont{Eriksson}},
  \bibinfo{author}{\bibfnamefont{J.~M.} \bibnamefont{Wills}}, \bibnamefont{and}
  \bibinfo{author}{\bibfnamefont{A.~M.} \bibnamefont{Boring}},
  \bibinfo{journal}{Phys. Rev. B} \textbf{\bibinfo{volume}{48}},
  \bibinfo{pages}{5844} (\bibinfo{year}{1993}).

\bibitem[{\citenamefont{Fast et~al.}(1995)\citenamefont{Fast, Wills, Johansson,
  and Eriksson}}]{FAS95}
\bibinfo{author}{\bibfnamefont{L.}~\bibnamefont{Fast}},
  \bibinfo{author}{\bibfnamefont{J.~M.} \bibnamefont{Wills}},
  \bibinfo{author}{\bibfnamefont{B.}~\bibnamefont{Johansson}},
  \bibnamefont{and} \bibinfo{author}{\bibfnamefont{O.}~\bibnamefont{Eriksson}},
  \bibinfo{journal}{Phys. Rev. B} \textbf{\bibinfo{volume}{51}},
  \bibinfo{pages}{17431} (\bibinfo{year}{1995}).

\bibitem[{\citenamefont{Phillips}(1959)}]{PHI59}
\bibinfo{author}{\bibfnamefont{N.~E.} \bibnamefont{Phillips}},
  \bibinfo{journal}{Phys. Rev.} \textbf{\bibinfo{volume}{114}},
  \bibinfo{pages}{676} (\bibinfo{year}{1959}).

\bibitem[{\citenamefont{Wolcott}(1957)}]{WOL57}
\bibinfo{author}{\bibfnamefont{N.~M.} \bibnamefont{Wolcott}},
  \bibinfo{journal}{Philos. Mag.} \textbf{\bibinfo{volume}{2}},
  \bibinfo{pages}{1246} (\bibinfo{year}{1957}).

\bibitem[{\citenamefont{Bechmann and Hearmon}(1966)}]{LANDOLT}
\bibinfo{author}{\bibfnamefont{R.}~\bibnamefont{Bechmann}} \bibnamefont{and}
  \bibinfo{author}{\bibfnamefont{R.~F.~S.} \bibnamefont{Hearmon}}, in
  \emph{\bibinfo{booktitle}{{L}andolt-{B}\"ornstein}}, edited by
  \bibinfo{editor}{\bibfnamefont{K.-H.} \bibnamefont{Hellwege}}
  \bibnamefont{and} \bibinfo{editor}{\bibfnamefont{A.~M.}
  \bibnamefont{Hellwege}} (\bibinfo{publisher}{Springer},
  \bibinfo{address}{Berlin}, \bibinfo{year}{1966}), vol.
  \bibinfo{volume}{{III}/1}.

\bibitem[{\citenamefont{Kamm and Alers}(1964)}]{KAM64}
\bibinfo{author}{\bibfnamefont{G.~N.} \bibnamefont{Kamm}} \bibnamefont{and}
  \bibinfo{author}{\bibfnamefont{G.~A.} \bibnamefont{Alers}},
  \bibinfo{journal}{J. Appl. Phys.} \textbf{\bibinfo{volume}{35}},
  \bibinfo{pages}{327} (\bibinfo{year}{1964}).

\bibitem[{\citenamefont{Nakamura and Kimura}(1991)}]{NAK91}
\bibinfo{author}{\bibfnamefont{M.}~\bibnamefont{Nakamura}} \bibnamefont{and}
  \bibinfo{author}{\bibfnamefont{K.}~\bibnamefont{Kimura}},
  \bibinfo{journal}{Journal of Material Science} \textbf{\bibinfo{volume}{26}},
  \bibinfo{pages}{2208} (\bibinfo{year}{1991}).

\bibitem[{\citenamefont{Fisher and Renken}(1964)}]{FIS64}
\bibinfo{author}{\bibfnamefont{E.~S.} \bibnamefont{Fisher}} \bibnamefont{and}
  \bibinfo{author}{\bibfnamefont{C.~J.} \bibnamefont{Renken}},
  \bibinfo{journal}{Phys. Rev.} \textbf{\bibinfo{volume}{135}},
  \bibinfo{pages}{482} (\bibinfo{year}{1964}).

\bibitem[{\citenamefont{Kikuchi}(1951)}]{KIK51}
\bibinfo{author}{\bibfnamefont{R.}~\bibnamefont{Kikuchi}},
  \bibinfo{journal}{Phys. Rev.} \textbf{\bibinfo{volume}{81}},
  \bibinfo{pages}{988} (\bibinfo{year}{1951}).

\bibitem[{\citenamefont{Saunders and Miodownik}(1998)}]{CALPHAD}
\bibinfo{author}{\bibfnamefont{N.}~\bibnamefont{Saunders}} \bibnamefont{and}
  \bibinfo{author}{\bibfnamefont{A.~P.} \bibnamefont{Miodownik}},
  \emph{\bibinfo{title}{{CALPHAD} - Calculation of Phase Diagrams - A
  Comprehensive Guide}} (\bibinfo{publisher}{Pergamon}, \bibinfo{year}{1998}).

\bibitem[{\citenamefont{Saunders}(1989)}]{SAU89}
\bibinfo{author}{\bibfnamefont{N.}~\bibnamefont{Saunders}},
  \bibinfo{journal}{Z. Metallkd.} \textbf{\bibinfo{volume}{80}},
  \bibinfo{pages}{894} (\bibinfo{year}{1989}).

\bibitem[{\citenamefont{Murray et~al.}(1992)\citenamefont{Murray, Peruzzi, and
  Abriata}}]{MUR92}
\bibinfo{author}{\bibfnamefont{J.}~\bibnamefont{Murray}},
  \bibinfo{author}{\bibfnamefont{A.}~\bibnamefont{Peruzzi}}, \bibnamefont{and}
  \bibinfo{author}{\bibfnamefont{J.~P.} \bibnamefont{Abriata}},
  \bibinfo{journal}{J. Phase Equilibria} \textbf{\bibinfo{volume}{13}},
  \bibinfo{pages}{277} (\bibinfo{year}{1992}).

\bibitem[{\citenamefont{Sigli et~al.}(1998)\citenamefont{Sigli, Maenner, Sztur,
  and Shahni}}]{SIG98}
\bibinfo{author}{\bibfnamefont{C.}~\bibnamefont{Sigli}},
  \bibinfo{author}{\bibfnamefont{L.}~\bibnamefont{Maenner}},
  \bibinfo{author}{\bibfnamefont{C.}~\bibnamefont{Sztur}}, \bibnamefont{and}
  \bibinfo{author}{\bibfnamefont{R.}~\bibnamefont{Shahni}}, in
  \emph{\bibinfo{booktitle}{Proceedings of ICAA-6, Aluminium Alloys}}
  (\bibinfo{year}{1998}), vol.~\bibinfo{volume}{1}, pp.
  \bibinfo{pages}{87--98}.

\bibitem[{\citenamefont{Asta et~al.}(1998)\citenamefont{Asta, Foiles, and
  Quong}}]{AST98}
\bibinfo{author}{\bibfnamefont{M.}~\bibnamefont{Asta}},
  \bibinfo{author}{\bibfnamefont{S.~M.} \bibnamefont{Foiles}},
  \bibnamefont{and} \bibinfo{author}{\bibfnamefont{A.~A.} \bibnamefont{Quong}},
  \bibinfo{journal}{Phys. Rev. B} \textbf{\bibinfo{volume}{57}},
  \bibinfo{pages}{11265} (\bibinfo{year}{1998}).

\end{thebibliography}

\end{document}